\documentclass[preprint,12pt]{elsarticle}

\usepackage{amsmath,amssymb,amsfonts}
\usepackage{url}
\usepackage{algorithmic}
\usepackage{graphicx}
\usepackage{textcomp}
\usepackage{hyperref}
\usepackage{multicol}
\usepackage{multirow}
\usepackage[table,xcdraw]{xcolor}
\usepackage{xcolor}
\usepackage{adjustbox}
\usepackage{ragged2e}
\usepackage{array}
\usepackage{mathtools}
\usepackage{rotating}
\usepackage{lipsum}
\usepackage{subfigure}
\usepackage[utf8]{inputenc}
\usepackage{booktabs}
\usepackage{adjustbox}
\usepackage{tabularx}
\usepackage[many]{tcolorbox}
\newtcolorbox{boxK}{
    sharpish corners, % better drop shadow
    boxrule = 0pt,
    toprule = 0.5pt, % top rule weight
    enhanced,
    fuzzy shadow = {0pt}{-2pt}{-0.5pt}{0.5pt}{black!35} 
}

\usepackage{geometry}
\journal{Journal of Systems and Software}

\begin{document}

\begin{frontmatter}

%% Title, authors and addresses

%% use the tnoteref command within \title for footnotes;
%% use the tnotetext command for theassociated footnote;
%% use the fnref command within \author or \address for footnotes;
%% use the fntext command for theassociated footnote;
%% use the corref command within \author for corresponding author footnotes;
%% use the cortext command for theassociated footnote;
%% use the ead command for the email address,
%% and the form \ead[url] for the home page:
%% \title{Title\tnoteref{label1}}
%% \tnotetext[label1]{}
%% \author{Name\corref{cor1}\fnref{label2}}
%% \ead{email address}
%% \ead[url]{home page}
%% \fntext[label2]{}
%% \cortext[cor1]{}
%% \affiliation{organization={},
%%             addressline={},
%%             city={},
%%             postcode={},
%%             state={},
%%             country={}}
%% \fntext[label3]{}

\title{Developer Challenges on Large Language Models: A Study of Stack Overflow and OpenAI Developer Forum Posts}

%% use optional labels to link authors explicitly to addresses:
%% \author[label1,label2]{}
%% \affiliation[label1]{organization={},
%%             addressline={},
%%             city={},
%%             postcode={},
%%             state={},
%%             country={}}
%%
%% \affiliation[label2]{organization={},
%%             addressline={},
%%             city={},
%%             postcode={},
%%             state={},
%%             country={}}

\author[inst1]{Khairul Alam}
\author[inst1]{Kartik Mittal}
% \author[inst1]{Banani Roy}
\author[inst1]{Banani Roy}
\author[inst1]{Chanchal Roy}           
\affiliation[inst1]{organization={Department of Computer Science, University of Saskatchewan},%Department and Organization
            addressline={110 Science Pl}, 
            city={Saskatoon},
            postcode={S7N 5C9}, 
            state={SK},
            country={Canada}}
\begin{abstract}
Large Language Models (LLMs) have gained widespread popularity due to their exceptional capabilities across various domains, including chatbots, healthcare, education, content generation, and automated support systems. Despite their transformative potential, developers encounter numerous challenges when implementing, fine-tuning, and integrating these models into real-world applications. This study investigates the challenges LLM developers face through an analysis of community interactions on Stack Overflow and the OpenAI Developer Forum. Using BERTopic modeling, we identify and categorize topics discussed by LLM developers. We also examine topics' popularity and difficulty. Our analysis yields nine evident challenges on Stack Overflow (e.g., LLM Ecosystem and Challenges, API Usage, LLM Training with Frameworks) and 17 on the OpenAI Developer Forum (e.g., API Usage and Error Handling, Fine-Tuning and Dataset Management, Prompt Engineering). Results indicate that developers frequently turn to Stack Overflow for implementation guidance, while OpenAI's forum is primarily used for troubleshooting. Additionally, on the OpenAI Developer Forum, API and functionality-related issues generated the most discussions, with many posts requiring multiple responses, highlighting the intricate nature of LLM challenges. We find that LLM-related queries often exhibit great difficulty, with a substantial percentage of unresolved posts (e.g., 79.03\% on Stack Overflow) and prolonged response times, particularly for complex topics like 'Llama Indexing and GPU Utilization' and 'Agents and Tool Interactions'. On the contrary, established fields such as Mobile development and Security enjoy faster resolution rates and greater community support. These findings emphasize the need for enhanced community support and tailored resources to assist LLM developers in addressing the complex, evolving challenges of this growing field. This study provides insights into areas where LLM developers encounter the most difficulty, guiding future research toward developing tools and techniques to better support the expanding community of LLM practitioners.
\end{abstract}

%%Graphical abstract
% \begin{graphicalabstract}
% \includegraphics{grabs}
% \end{graphicalabstract}

% %%Research highlights
% \begin{highlights}
% \item Research highlight 1
% \item Research highlight 2
% \end{highlights}


\end{frontmatter}

%% \linenumbers

%% main text
\section{Introduction}
The field of large language models (LLMs) experienced a groundbreaking transformation with the advent of transformer architecture, introduced by Vaswani et al. in their influential paper, \emph{"Attention Is All You Need"} \cite{vaswani2017attention}. This architecture, which is unique for avoiding the use of recurrence, instead relies on a self-attention mechanism. This allows transformers to weigh input tokens throughout a sequence, effectively capturing the context of words, no matter how far apart they are positioned. Building on transformer architecture, groundbreaking models like BERT and the GPT series—especially GPT-3 and GPT-4—have significantly advanced language processing. BERT introduced new methods for understanding language context, while GPT models, with their autoregressive design, excel in tasks such as language generation, translation, and summarization \cite{brown2020language}. These advancements have made language models vital tools in natural language processing, enabling diverse real-world applications. LLMs have sparked significant interest across both academic and industry domains \cite{zhao2023survey, wei2022emergent, wang2023cmb}. Their impressive performance has led to excitement around their potential to embody Artificial General Intelligence (AGI) in this era \cite{bubeck2023sparks}. LLMs have the capacity to solve diverse tasks, compared with prior models, which are confined to solving particular tasks \cite{chang2024survey}. Their versatility in handling both general and domain-specific language tasks has made them invaluable for users with critical information needs, such as students and researchers. The wide applicability of LLMs highlights the crucial need to ensure their safety and reliability, particularly in sensitive fields like finance, healthcare, and aerospace.

LLMs are being used in many fields to improve tasks that involve understanding, generating, or interpreting language. In healthcare, for instance, they help summarize medical records, diagnose, and provide insights for patient care \cite{nazi2024large}. In finance, LLMs are used in customer service chatbots, fraud detection, and market trend analysis \cite{zhao2024revolutionizing}. Education has also been gained from LLMs, with applications in personalized tutoring, grading support, and language translation \cite{kasneci2023chatgpt}. They are even used in creative fields like writing, content creation, and art, showing their flexibility and adaptability \cite{floridi2020gpt}. As LLMs continue to advance, they are making a significant impact by supporting both specialized and general tasks, transforming workflows, and aiding in decision-making across various industries.

LLMs are becoming valuable tools in software engineering, helping with everything from writing code to reviewing and documenting it. Models like Codex and ChatGPT assist developers by suggesting code snippets, completing functions, and even filling in code blocks, which boosts productivity and saves time \cite{zhang2023planning, liu2024your}. LLMs also play a role in code review and quality assurance by spotting bugs, improving code, and enforcing good practices, leading to more reliable software \cite{yu2024security}. Additionally, they’re useful for generating and updating documentation, such as code comments, user guides, and API documentation, making software easier to understand and manage \cite{nam2024using, su2023hotgpt}. LLMs can even translate code between programming languages, helping with cross-platform development and improving collaboration across teams with diverse technical skills \cite{pan2023understanding, yang2024exploring}. These applications show how LLMs are transforming software engineering by enabling faster, more accurate, and collaborative work.

With the growing popularity of ChatGPT by OpenAI and the fast pace of advancements in LLM technology, more and more developers are beginning to use these tools in their work. However, developers often face various challenges when working with LLMs, such as setting up their environment, managing API calls, configuring parameters, and handling errors. Building plugins or applications with these models can be incredibly challenging for those new to AI and LLMs. The process may require integrating multiple tools or applications, addressing security concerns, and optimizing performance. Additionally, maintaining data privacy and security is critical; developers need to follow regulations and put proper security practices in place to protect user data.

Compared to traditional software development, LLM development brings unique challenges that require specialized skills and expertise. LLM developers often need to automate complex task processing, manage uncertainty in model behavior, work with diverse data types, and interpret model outputs effectively. These challenges highlight the distinct nature of LLM development, where understanding nuances in language data and model predictions is crucial. Additionally, LLMs often require careful tuning and monitoring to ensure reliable performance, accuracy, and ethical use, adding layers of complexity beyond standard software projects. Recognizing these challenges is essential for understanding the evolving landscape of LLM development and supporting developers as they navigate this dynamic field.

In this paper, we conduct an empirical study to rigorously investigate the challenges faced by developers working with LLMs, aiming to understand their needs and hurdles. Our objective is to identify specific issues that developers encounter and highlight areas where support and improvements are needed. Developers frequently turn to Stack Overflow to ask questions, find solutions, explore new technologies, validate best practices, and connect with a vast community \cite{mamykina2011design}. Stack Overflow has become an essential resource for LLM developers \cite{son2023trend}, fostering shared learning and collaboration that drive open-source projects and collective innovation. As the world's largest developer community, its importance goes beyond troubleshooting; Stack Overflow reflects trends and shifts in the technology landscape. In the realm of LLMs, it provides valuable insights into real-world applications and the specific challenges developers face. Research by Barua et al. \cite{barua2014developers} sheds light on developer interactions on Stack Overflow, revealing the platform's role in addressing emerging technology challenges. Despite the growing interest in LLMs and the critical role of platforms like Stack Overflow in showcasing technology trends, in-depth analyses of LLM-specific developer challenges remain scarce. To address this, we extract LLM-related 8,593 unique posts from Stack Overflow. Additionally, we analyzed 26,474 unique questions from the OpenAI Developer Forum. This forum serves as a collaborative hub for LLM developers working with OpenAI technologies, offering support on topics such as API integration, plugin development, and best practices. 

Utilizing both Stack Overflow and the OpenAI Developer Forum is essential to capture a comprehensive view of developer challenges with LLMs. Stack Overflow provides a broad perspective, showcasing real-world issues and common obstacles faced across varied LLM applications. In contrast, the OpenAI Developer Forum offers focused insights specific to OpenAI's LLM technologies, revealing challenges unique to OpenAI's ecosystem. Together, these platforms provide complementary data that enhances understanding of both general and platform-specific LLM development hurdles. Thus, we utilize both platforms for our study.

Using a combination of manual analysis and automated topic modeling, we examine these Stack Overflow and OpenAI Developer Forum posts to uncover developers' specific challenges. In particular, our study aims to answer the following three research questions:
\begin{itemize}
    \item \textbf{RQ1: What topics are Large Language Model developers are asking about?} Posts related to large language models (LLMs) on Stack Overflow and the OpenAI Developer Forum reveal various challenges developers face when working with LLMs. To identify these challenges, we use BERTtopic modeling to pull out specific topics from these data. From Stack Overflow posts, we uncover nine topics, covering both traditional software engineering issues (e.g., API Usage, Environment Management) and LLM-focused topics (e.g.,  LLM Training with Frameworks, Langchain Development and Error Handling). Similarly, the analysis of OpenAI Developer Forum data yields 17 topics, again including traditional software topics (e.g., API Usage and Error Handling, File Management and Retrieval) and LLM-specific topics (e.g., Fine-Tuning and Dataset Management, Prompt Engineering). By analyzing the challenges discussed on Stack Overflow and the OpenAI Developer Forum, we aim to answer \emph{RQ1}, identifying areas in LLM development where developers most frequently seek help, thus highlighting the aspects of LLM development that often prove challenging.
    \item \textbf{RQ2: What types of questions are  Large Language Model developers are asking?} Developers ask a variety of questions to address different challenges in LLM development. To answer RQ2, we use a method similar to previous studies on categorizing Stack Overflow posts \cite{rosen2016mobile, abdellatif2020challenges, treude2011programmers}. We select a statistically significant sample of posts for each topic and classify them into types like \emph{How, What, Why} and \emph{Other}. For the OpenAI Developer Forum data, we utilize our expertise and ChatGPT-4 to identify question types related to LLM development. Based on that, we establish eight main types: \emph{Model Development and Deployment, Evaluation and Optimization, Ethical and Best Practices, Troubleshooting, Feature Requests and Improvements, Guidance and Validation, Maintenance and Ongoing Support}, and \emph{Other} for each topic. This analysis uses a statistically significant sample size (95\% confidence with a 5\% margin of error).

    On Stack Overflow, developers frequently seek guidance on specific implementation steps, API usage examples, and troubleshooting—indicating a high prevalence of \emph{How} questions, similar to trends in other fields like chatbot and mobile development. This highlights developers' need for practical, step-by-step instructions. From OpenAI data, \emph{Troubleshooting} stands out as the most common question type, reflecting developers' challenges in using these models effectively. \emph{Feature Requests and Improvements} and \emph{Guidance and Validation} also represent a notable share of posts, indicating a demand for support in refining and enhancing LLM capabilities.
    
    \item \textbf{RQ3: To what extent do developers perceive the revealed challenges in terms of difficulties?} To assess the difficulty of Stack Overflow topics, we use two metrics: the percentage of unresolved posts and the median time to resolve them. This approach aligns with approaches in similar studies \cite{rosen2016mobile, bagherzadeh2019going, yang2016security, li2021understanding, scoccia2021challenges, abdellatif2020challenges}. For OpenAI Developer Forum data, we look at the number of replies to LLM development-related questions, as we can't measure exact response times, and accepted answers are not marked in the forum. This approach is also used in \cite{chen2024empirical}. We find that over 51\% of questions have fewer than three replies, suggesting challenges in those topics. The most difficult topics appear to be \emph{API Usage and Error Handling}, \emph{LLM Functionalities}, \emph{Data Preparation and Structured Analysis}, and \emph{Function Parameters and Callback Handling}, as these have a high concentration of posts with few replies, showing a lack of answers. For Stack Overflow, the topic \emph{Agents and Tool Interactions} has the highest difficulty, with 90.63\% of posts unresolved and a median response time of 36.51 hours. Similarly, \emph{Llama Indexing and GPU Utilization} is challenging, with 85.07\% of posts unresolved and the longest median response time of 55.76 hours.
\end{itemize}

\textbf{Paper Organization: }The remainder of this paper is structured as follows: In Section \ref{background-and-related-work}, we briefly discuss LLMs, Topic Modeling, and related work. Moving on to Section \ref{study-design}, we detail our study's methodology. Our findings are presented in Section \ref{case-study-results}, and Section \ref{discussion-implications} illustrates the implications. Section \ref{threats-to-validity} addresses potential threats to the validity of our results. Finally, Section \ref{conclusion} concludes the paper, highlighting directions for future research.

\section{Background and Related Work} \label{background-and-related-work}
This study aims to understand the challenges faced by LLMs developers by analyzing Stack Overflow and the OpenAI Developer Forum posts. In this section, we present an overview of the background and relevant research. First, we explore the background and related work on LLMs, followed by a review of previous studies that employed topic modeling techniques to uncover insights into developer perspectives.
\subsection{Large Language Models (LLMs)}
Language modeling has been studied for over two decades and has evolved from statistical methods to neural models. Pre-trained language models (PLMs) emerged by training transformer-based models on vast datasets, proving effective in various NLP tasks. Researchers found that scaling up these models enhances their capabilities \cite{zhao2023survey}. When the parameter scale exceeds a certain level (tens or hundreds of billions of parameters), they perform better and display unique abilities like contextual learning, which smaller models lack. To distinguish the language models in different parameter scales, the research community has introduced the term large language models (LLMs) for the PLMs of significant size \cite{zhao2023survey, kaplan2020scaling}. These LLMs have seen rapid development in academia and industry, exemplified by innovations like ChatGPT, fundamentally impacting AI's development and applications.

Large Language Models (LLMs) are typically transformer-based language models with hundreds of billions of parameters, trained on vast amounts of text data to understand, generate, and process human language \cite{shanahan2024talking}. Examples include OpenAI's GPT models (e.g., GPT-3 \cite{brown2020language}, GPT-4 \cite{achiam2023gpt}), Google's PaLM \cite{chowdhery2023palm}, and Meta's LLaMA \cite{touvron2023llama}. These models are competent in understanding natural language and handling complex tasks.
\subsection{Prior Studies on LLMs}
The rapid success of LLMs across a wide range of applications has spurred a significant increase in LLM-focused research contributions. This influx is evident in various survey studies, which aim to consolidate and evaluate the current LLM literature \cite{zhao2023survey, naseem2021comprehensive, min2023recent, zhou2023comprehensive}. Numerous studies have delved into understanding and expanding the capabilities of LLMs, each highlighting unique domains of application and emerging challenges. For instance, Kasneci et al. \cite{kasneci2023chatgpt} explore the opportunities LLMs bring to the field of education, examining how they can enhance learning experiences and support both students and educators. In the medical domain, Thirunavukarasu et al. \cite{thirunavukarasu2023large} investigate the potential uses of LLMs for clinical applications and medical research, showcasing their promise in diagnostics, patient management, and bioinformatics. A broader evaluation of LLM capabilities is presented by Chang et al. \cite{chang2024survey}, where they conducted a comprehensive survey of LLM functionalities and limitations across different fields.

The emergent abilities of LLMs, which refer to capabilities that develop as models scale in size and complexity, are discussed by Wei et al. \cite{wei2022emergent}, offering insight into how these models adapt and evolve with more extensive training. Similarly, Huang et al. \cite{huang2022large} investigate the self-improvement mechanisms in LLMs, discussing how they can be fine-tuned or adapted for continual learning. Furthermore, Kaddour et al. \cite{kaddour2023challenges} identify critical challenges in LLM deployment, such as bias, interpretability, and scalability, while outlining their applications in real-world scenarios. In machine translation, Brants et al. \cite{brants2007large} examine the challenges LLMs face, including linguistic diversity and context retention, illustrating the complexities in achieving reliable language translation at scale. Additionally, the use of LLMs for recommendation systems is discussed by Wu et al. \cite{wu2024survey}, where their role in personalizing content and improving user engagement is evaluated.

LLMs have shown great potential in supporting various software engineering tasks. By being trained on extensive datasets, LLMs excel at understanding, processing, and generating code across multiple programming languages. LLMs are widely used in studies related to code generation \cite{zhang2023planning, vaithilingam2022expectation, jiang2024self, poesia2022synchromesh}, where they assist in refining and improving the code creation process, making it easier to produce high-quality code efficiently \cite{liu2024your, liu2024refining}. In addition, LLMs have shown promise in automatic program repair by detecting and fixing bugs \cite{weng2023automatic, bouzenia2024repairagent, jin2023inferfix}, reducing the time and effort needed for manual debugging \cite{xia2023keep}. They are also used for managing and categorizing bug reports \cite{du2024llm}, evaluating bug report summaries \cite{kumar2024llms}, and reproducing bug reports \cite{kang2023large}, which enhances the overall efficiency of software maintenance \cite{li2024enhancing, zhang2023cupid}. LLMs contribute to the automatic testing process, enabling broader test coverage and faster issue identification \cite{liu2024testing, pizzorno2024coverup, chen2024chatunitest}. They also play a role in identifying potential security vulnerabilities in code \cite{lu2024grace, akuthota2023vulnerability}, aiding in the development of more secure software applications \cite{chen2023chatgpt, fu2023chatgpt}. Additionally, LLMs support code summarization \cite{ahmed2022few, ahmed2024automatic}, which helps developers understand and maintain codebases more effectively. The expanding body of research reflects an ongoing effort to enhance LLM capabilities, address ethical and operational challenges, and discover novel applications across diverse fields. Researchers continue to push the boundaries of what LLMs can achieve, contributing to a rich knowledge landscape that supports these powerful models' evolution and deployment.

Given the rapid expansion of LLMs across various domains—particularly in software engineering—identifying and understanding the challenges LLM developers face is crucial. These challenges not only affect the efficacy of LLMs in real-world applications but also impact their scalability, security, and ethical deployment. Despite the importance, research into developer challenges surrounding LLMs remains limited. 
Existing studies provide valuable insights but often lack the depth needed to fully capture the diverse difficulties encountered by LLM developers. For example, Chen et al. \cite{chen2024empirical} constructed a taxonomy of challenges using a limited sample of 2,364 questions exclusively from OpenAI data. Similarly, Son et al. analyzed Stack Overflow data to track trends in LLM-related issues, while Ullah et al. \cite{ullah2024challenges} discussed significant issues like hallucination, toxicity, bias, fairness, and privacy concerns. Patil et al. \cite{patil2024review} extended this exploration by examining toxic content, hallucination, cost, environmental impact, and biases within LLMs. However, these studies mostly focus on the challenges of using LLMs from a user’s perspective rather than directly analyzing the challenges that LLM developers face.

To address this gap, our study aims to comprehensively identify the challenges faced by LLM developers by examining two key sources of information: Stack Overflow, the most popular technical Q\&A platform, and the OpenAI Developer Forum, a dedicated community for LLM developers. By analyzing both platforms, we aim to capture a holistic view of developer pain points across different stages of LLM deployment, including training, fine-tuning, and ethical considerations. This approach will offer a broader and more nuanced understanding of the difficulties LLM developers encounter, informing future research and improvements in LLM development processes.
\subsection{Topic Modeling}
Topic modeling is a statistical technique used to uncover the underlying structure and main themes within a large set of documents, helping to organize, search, and better understand the content \cite{jelodar2019latent}. It is applied in various text analysis tasks \cite{isoaho2021topic}, including information retrieval \cite{yi2009comparative, wei2007topic}, document clustering \cite{yau2014clustering, xie2013integrating}, recommendation systems \cite{luostarinen2013using, choi2015improving, bergamaschi2015comparing}, content summarization \cite{belwal2023extractive}, and language translation \cite{eidelman2012topic}. Topic modeling has been extensively researched, with numerous studies published across fields such as software engineering, social media analysis, e-commerce and product analysis, medicine, and linguistics. The software engineering community, in particular, has seen significant use of topic modeling \cite{asuncion2010software, silva2021topic}, especially in areas like mining software repositories \cite{zhai2011constrained, chen2012explaining, chen2016survey, thomas2011mining, thomas2011modeling, tian2009using}, source code analysis \cite{gethers2010using, linstead2008application, lukins2010bug, savage2010topic, tian2009using}, spam detection \cite{10527342, li2013topicspam}, and recommendation systems \cite{cheng2016effective, kim2014twilite, lu2015twitter, zhao2016personalized, zoghbi2016latent}.

Various methods, including BERTopic \cite{grootendorst2022bertopic}, Latent Dirichlet Allocation (LDA) \cite{blei2003latent}, Correlated Topic Model (CTM) \cite{blei2006correlated}, Dynamic Topic Models \cite{blei2006dynamic}, Non-Negative Matrix Factorization (NMF) \cite{lee2000algorithms}, Word Embedding-Based Models \cite{blei2006dynamic}, and Biterm Topic Model (BTM) \cite{yan2013biterm}, have each contributed significantly to advancements in topic modeling.

Previous studies have commonly used the LDA algorithm \cite{blei2003latent} or its variants for topic extraction. However, recent research has increasingly adopted BERTopic \cite{grootendorst2022bertopic} for similar analyses, including the study of scientific papers \cite{10285737}, spam detection \cite{10527342}, research trends on language models \cite{10487248, doi2024topic}, natural language processing \cite{9854488}, and machine learning integration \cite{atzeni2022systematic}. While LDA has served as a foundational method in topic modeling, it has notable limitations. For example, LDA requires the number of topics to be set in advance, which can be challenging to determine and often requires manual adjustments. Its reliance on a bag-of-words representation disregards word order and context, potentially leading to a loss of key semantic information. Additionally, as datasets grow in size or as the number of topics increases, LDA’s computational demands escalate, reducing its scalability. The algorithm is also prone to producing incoherent topics, especially in noisy or extensive datasets, and is sensitive to hyperparameter configurations, demanding extensive experimentation.

In contrast, BERTopic \cite{grootendorst2022bertopic} overcomes these limitations with advanced capabilities. It automatically determines the number of topics, captures contextual information through transformer-based models like BERT, and handles larger datasets more effectively. BERTopic also incorporates hierarchical clustering and dynamic topic modeling, allowing it to adapt to data changes and generate more granular, interpretable topics. These strengths make BERTopic highly suitable for applications requiring a deep semantic understanding. Given the concise and technical nature of data like Stack Overflow and OpenAI Developer Forum posts, BERTopic is particularly advantageous due to its capacity to capture nuanced semantics and contextually relevant topics. Consequently, this study plans to utilize BERTopic \cite{grootendorst2022bertopic} to leverage these capabilities for improved topic extraction.

\subsection{Topic Analysis of Technical Q\&As}
Topic modeling plays a vital role in analyzing technical Q\&A data by enabling the efficient categorization and organization of vast collections of questions and answers \cite{chen2023user}. By uncovering underlying themes and topics within the content, topic modeling helps structure and index data, making it more accessible and easier for users to locate relevant information. This process enhances knowledge discovery and supports users in navigating large, complex datasets effectively \cite{daud2010knowledge}.

For instance, in forums like Stack Overflow and the OpenAI Developer Forum, where developers can ask and answer questions, share insights, and discuss challenges, topic modeling can cluster similar questions, facilitating faster retrieval of solutions and minimizing redundancy \cite{nie2017data}. Additionally, topic modeling can track trending topics, monitor user interests, and identify common issues or knowledge gaps, which is beneficial for both users seeking information and developers looking to enhance their products or services \cite{fiscus2002topic}. This capability allows forums to stay updated with user needs, ensuring that both frequent and emerging issues are addressed effectively.

Topic modeling has been extensively applied in research to understand the themes of general Stack Overflow posts \cite{allamanis2013and} and track topic trends over time \cite{uddin2021empirical, barbosa2020software, wang2013empirical, chen2019modeling, zou2015non}. Prior work has also leveraged topic modeling in specific development domains, including mobile applications \cite{rosen2016mobile}, machine learning \cite{alshangiti2019developing}, concurrency \cite{ahmed2018concurrency}, security \cite{yang2016security}, DevOps \cite{10371495}, chatbot development \cite{abdellatif2020challenges}, database technology classification \cite{9509047}, and understanding nonfunctional requirements \cite{zou2017towards}. However, topic modeling analysis specifically for the OpenAI Developer Forum remains largely unexplored. To the best of our knowledge, a comprehensive topic analysis of LLM-related discussions on both Stack Overflow and the OpenAI Developer Forum has yet to be conducted despite LLMs representing a rapidly advancing area in software engineering. Recognizing the value in identifying the key topics surrounding LLMs, we utilize BERTopic \cite{grootendorst2022bertopic} due to its robust contextual understanding and high accuracy. This approach allows us to extract and analyze topics more precisely, providing insights into the current challenges, trends, and areas of interest in LLM development.

\begin{figure}[htbp]
\centering
  \vspace{-0.7em}
  \includegraphics[width=\textwidth]{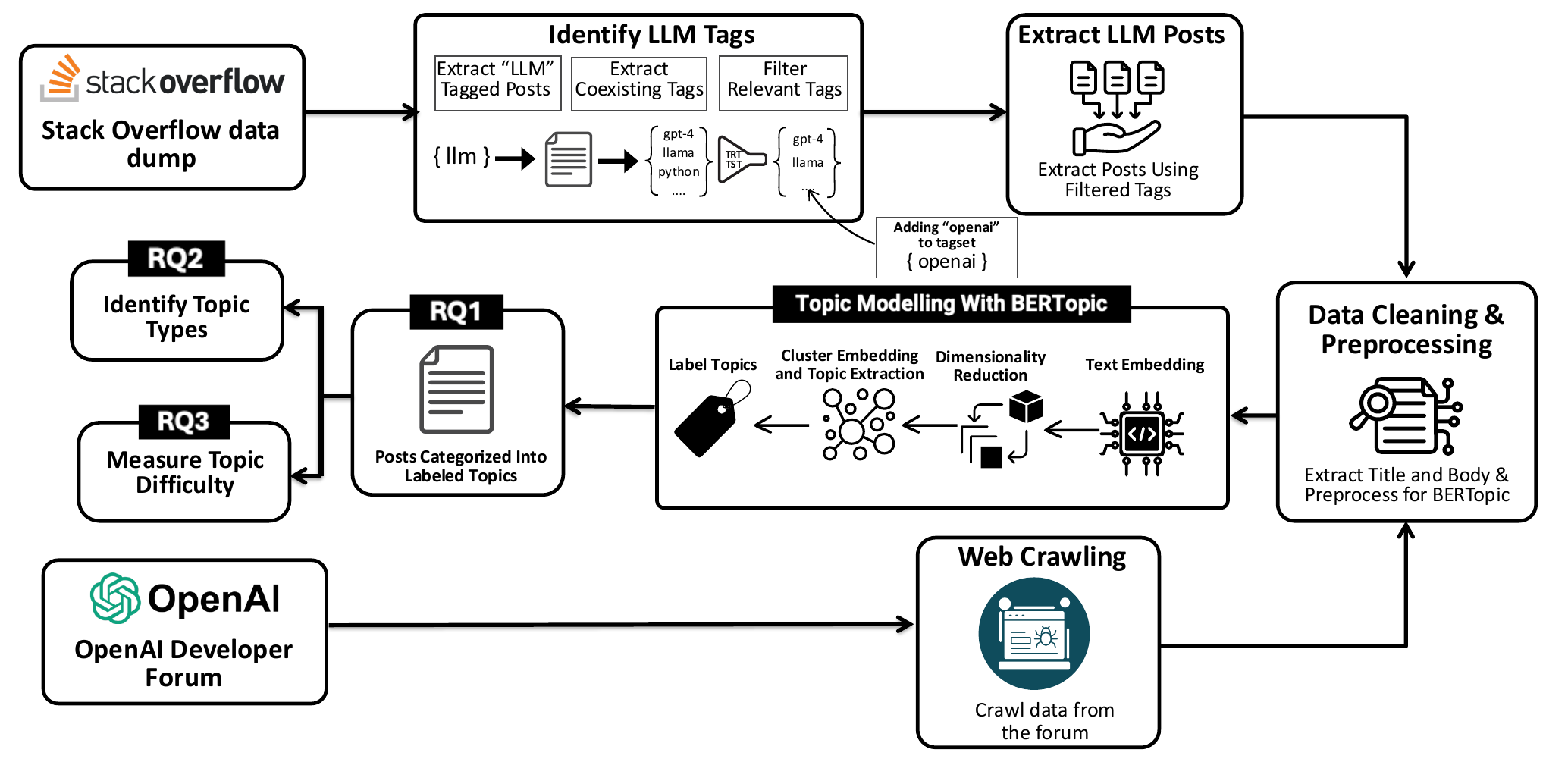}
  \caption{Overview of the methodology of our study}
  \label{fig:studydesign}
  \vspace{-2.2em}
\end{figure}

\section{Study Design} \label{study-design}
To investigate the challenges faced by developers working with LLMs, we collect and analyze a dataset comprising 8,593 posts from Stack Overflow and 26,474 posts from the OpenAI Developer Forum \cite{openaidevforum}, an official platform provided by OpenAI for developers engaging with their LLM products. Our study follows established methodologies from prior research \cite{ahmed2018concurrency, alshangiti2019developing, bagherzadeh2019going, barua2014developers, rosen2016mobile, li2021understanding, han2020programmers, abdellatif2020challenges, lou2020understanding, yang2016security, zhang2019empirical} to systematically mine and process data from both sources. These references provided critical insights and techniques for effectively extracting and analyzing data from developer communities. Figure \ref{fig:studydesign} presents an overview of our study methodology.

The main objective of this study is to investigate the topics of inquiries made by developers working with LLMs. To achieve this, we analyze developers' discussions on Stack Overflow, a platform that offers a rich dataset of questions, answers, and associated metadata (e.g., accepted answers). Stack Overflow has been widely used in similar studies across various domains, such as chatbots \cite{abdellatif2020challenges}, cryptography APIs \cite{nadi2016jumping}, deep learning \cite{han2020programmers}, concurrency  \cite{ahmed2018concurrency}, and quantum software engineering \cite{li2021understanding}. However, while the platform provides structured data, it lacks fine-grained information specific to LLM-related content. Consequently, our first step is to gather posts related to LLMs, identify the topics of the posts, and subsequently perform our analysis.

OpenAI has been a leading force in the research and development of LLMs, particularly through its GPT (Generative Pre-trained Transformer) models, which showcase the advanced capabilities of LLMs in various tasks such as text generation, contextual understanding, summarization, translation, question-answering, and code generation. OpenAI’s GPT models, evolving from GPT-2 to the more sophisticated GPT-4, have set benchmarks in the field of natural language processing, supporting a broad spectrum of applications, including customer service, content creation, and software development. In addition to its technical advancements, OpenAI has emphasized the ethical development and responsible deployment of LLMs, addressing societal concerns such as bias, misinformation, and AI safety. This leadership underscores the significance of OpenAI's role not only in advancing LLM technology but also in navigating its societal impact.

Thus, we feel the necessity to explore the challenges developers face when working with LLMs. By examining questions and discussions on the OpenAI Developer Forum alongside data from Stack Overflow, we aim to identify and analyze the common issues and obstacles developers encounter. This study focuses on extracting and comparing developer challenges from both platforms to provide a comprehensive understanding of the difficulties LLM developers face. The following steps outline the data collection methods and analytical approaches used in this investigation.

\subsection{Stack Overflow} \label{so-topics}
\textbf{Step 1: Download posts containing large-language-model tag.} We gather posts from Stack Overflow using the Stack Exchange Data Explorer \cite{stackexchangeforum} up until September 14, 2024, filtered by the tag \emph{large-language-model}. The initial dataset consists of 1,693 posts tagged with \emph{large-language-model}.
\newline
\textbf{Step 2: Identify large-language-model tags.} Stack Overflow hosts posts on a wide range of software development topics, such as Java, Python, security, and blockchain. Authors typically use popular tags (e.g., large-language-model, chatbot, and web) to enhance the visibility of their posts and increase the likelihood of receiving answers \cite{barua2014developers}. To identify the most relevant tags related to LLMs, we follow the approach used in previous studies \cite{abdellatif2020challenges, bagherzadeh2019going, rosen2016mobile}, and create a tag set using the following procedure.

First, we retrieve all posts with the 'large-language-model' tag, resulting in a dataset of 1,693 (2024-09-14) posts. To minimize the risk of introducing noise, we avoid adding any additional tags at this initial stage, as these posts will be used to identify other tags related to LLMs. Next, we extract all tags that co-occur with the 'large-language-model' tag from these posts. To expand the set of LLM-related tags, we apply two heuristic metrics from prior research \cite{abdellatif2020challenges, rosen2016mobile, wan2019programmers}. The first metric is the Tag \emph{Relevance Threshold (TRT)}, which measures how closely a tag is related to large-language-model-tagged posts. This is calculated as the ratio of posts containing both the 'large-language-model' tag and the specific tag to the total number of posts for that tag. Specifically, the TRT is measured using the following equation. \[
TRT_{\text{tag}} = \frac{\text{No. of LLMs posts for the tag}}{\text{Total No. of posts for the tag}}
\] 

We get distinct tags and their occurrence count from the downloaded 1693 posts. Then, we count the occurrence of all the tags for the whole dataset using the \emph{tags} table of the Stack Overflow database. Then we calculate the TRT. For example, the tag 'langchain' has a TRT of 24.98\%, indicating that 24.98\% of posts tagged with 'langchain' are also tagged with 'large-language-model'. By utilizing the TRT, we can effectively remove irrelevant tags from our tag set.

However, some tags with a small number of posts (e.g., the 'mamba-ssm' tag, which has only 3 posts) may have a high TRT (33.3\%) simply because a single post is related to LLMs. This can introduce less significant tags into our analysis. To address this, we incorporate a second metric, the \emph{Tag Significance Threshold (TST)}, which measures the prominence of a tag within LLM-related posts. This metric is determined by comparing the total number of LLM-related posts for a specific tag to the total number of LLM-related posts for the most popular tag (the 'large-language-model' tag with 1,693 posts). The \emph{TST} is calculated as the ratio of LLM-related posts for a given tag to the number of posts for the most popular tag. The equation for calculating \emph{TST} is shown below. \[TST_{\text{tag}} = \frac{\text{No. of LLMs posts for the tag}}{\text{No. of LLMs posts of the most popular tag}}
\]
For example, the 'langchain' tag has a TST of 32.33\%, meaning that the number of posts tagged with both 'langchain' and 'large-language-model' is equivalent to 32.33\% of the total number of LLM-related posts tagged with 'large-language-model'.

\begin{table}[ht]
\centering
\caption{The tag set used to identify the LLMs-related posts. The TRT and TST are expressed in percentages.}
\label{tab:tagststtrt}
\begin{tabular}{|>{\centering\arraybackslash}m{4cm}|>{\raggedleft\arraybackslash}m{1.5cm}|>{\raggedleft\arraybackslash}m{1.5cm}|>{\centering\arraybackslash}m{0.0001cm}|>{\centering\arraybackslash}m{3cm}|>{\raggedleft\arraybackslash}m{1.5cm}|>{\raggedleft\arraybackslash}m{1.5cm}|}
\toprule
\textbf{Tag Name} & \textbf{TRT} & \textbf{TST} &  & \textbf{Tag Name} & \textbf{TRT} & \textbf{TST} \\ 
\midrule
large-language-model & 100 & 100 & & ctransformers & 56.25 & 0.5316 \\ 
langchain & 24.98 & 32.13 & & huggingface-hub & 13.04 & 0.5316 \\ 
llama & 41.09 & 11.16 & & vllm & 32.14 & 0.5316 \\ 
huggingface & 13.94 & 8.51 & & langgraph & 17.50 & 0.4135 \\ 
llama-index & 25.76 & 6.02 & & lm-studio & 87.50 & 0.4135 \\ 
py-langchain & 16.46 & 3.96 & & claude & 16.22 & 0.3544 \\ 
chromadb & 19.26 & 3.37 & & dspy & 46.15 & 0.3544 \\ 
retrieval-augmented-generation & 40.58 & 3.31 & & anthropic & 33.33 & 0.2953 \\ 
fine-tuning & 17.91 & 3.13 & & flowise & 20.83 & 0.2953 \\ 
ollama & 26.74 & 2.95 & & gemma & 28.57 & 0.2363 \\ 
mistral-7b & 56.63 & 2.78 & & langsmith & 26.67 & 0.2363 \\ 
peft & 48.57 & 2.01 & & semantic-kernel & 10.26 & 0.2363 \\ 
rag & 38.67 & 1.71 & & text-chunking & 10.81 & 0.2363 \\ 
llama-cpp-python & 39.06 & 1.48 & & chainlit & 17.65 & 0.1772 \\ 
pinecone & 16.79 & 1.30 & & databricks-dolly & 60.00 & 0.1772 \\ 
openaiembeddings & 17.70 & 1.18 & & few-shot-learning & 14.29 & 0.1772 \\ 
llamacpp & 31.15 & 1.12 & & llama3.1 & 25.00 & 0.1772 \\ 
gpt-4 & 12.59 & 1.06 & & graphrag & 33.33 & 0.1181 \\ 
langchain-agents & 34.88 & 0.89 & & h2ogpt & 50.00 & 0.1181 \\ 
google-generativeai & 15.05 & 0.83 & & huggingface-evaluate & 15.38 & 0.1181 \\ 
gpt4all & 27.45 & 0.83 & & langchain4j & 11.11 & 0.1181 \\ 
llama3 & 30.43 & 0.83 & & mixtral-8x7b & 33.33 & 0.1181 \\ 
text-generation & 35.90 & 0.83 & & privategpt & 16.67 & 0.1181 \\ 
amazon-bedrock & 12.15 & 0.77 & & retrievalqa & 33.33 & 0.1181 \\ 
huggingface-trainer & 12.50 & 0.77 & & safe-tensors & 16.67 & 0.1181 \\ 
\bottomrule
\end{tabular}
\end{table}

We consider a tag to be significant and relevant to LLM-related posts if both its TRT and TST exceed certain thresholds. To determine these thresholds, three individuals with varying levels of LLM development experience independently examine tags with different TRT and TST values. For each tag, they review a randomly selected sample of posts to assess when the tag became less relevant or specific to LLMs. This process aims to identify the most appropriate TRT and TST thresholds, a method employed in several previous studies \cite{abdellatif2020challenges, bagherzadeh2019going, rosen2016mobile}. The goal is to select tags that are relevant to LLMs while minimizing noise in the dataset.

After evaluating the tags, the three individuals discussed their findings to reach a consensus on the optimal TRT and TST values. They independently assessed the thresholds that yielded the best results and deliberated to agree on a final decision. We find that tags with a TRT higher than 11\% and a TST greater than 0.11\% yield an appropriate balance between including more LLM-related posts (resulting in a more representative dataset) and filtering out posts unrelated to LLMs (reducing noise). Importantly, our thresholds align with those used in previous studies that adopted a similar approach. Finally, we apply the selected TRT and TST thresholds to define our tag set. Table \ref{tab:tagststtrt} presents the tags included in our tag set along with their corresponding TRT and TST values. In addition to our primary tag set, we also include the 'openai' tag for data extraction, as OpenAI plays a pivotal role in the development of LLMs. Given OpenAI's extensive work in advancing LLM technology, including models like GPT-3 and GPT-4, incorporating this tag helps ensure that we capture relevant discussions and developments related to cutting-edge LLM innovations. This broader scope allows us to analyze insights from the forefront of LLM research, making our dataset more comprehensive and aligned with the latest trends in LLM advancements.
\newline
\textbf{Step 3: Extract large-language-model posts.} After obtaining the LLM-related tag set, we use these tags to extract the posts that will form the basis of our LLM dataset for this study. We gather this corpus by querying all posts on Stack Overflow that are tagged with any of the tags in our set. This process resulted in a dataset containing 8,593 unique LLM posts along with their respective metadata.
% Our dataset can be found in the replication package.
\newline
\textbf{Step 4: Preprocessing large-language-model posts. }We begin by filtering out irrelevant information before applying topic modeling techniques. While the post title offers a concise summary of the question, the body provides essential context and details that aid in accurately identifying the discussed topic. In this analysis, we consider both the post title and body. However, the body content can introduce noise, so we clean it by removing quotes, HTML tags, links, and code snippets using regular expressions. We also eliminate stopwords—common words in the English language such as 'how', 'can', and 'at'—which do not significantly impact the meaning of a sentence and may introduce bias. For this, we rely on the NLTK stopwords list \cite{bird2006nltk}. Additionally, we apply lemmatization, a process that reduces words to their base or canonical forms (for instance, "scale" being the lemma of "scaling"), considering their linguistic context. The outcome is a processed dataset ready for input into the topic modeling phase.
% All preprocessing steps, along with the processed data, are provided in the replication package.
\newline
\textbf{Step 5: Identify large-language-model topics: }BERTopic begins by converting input documents into numerical representations, which is a crucial step in topic modeling. Several methods exist for this transformation, but \emph{SentenceTransformers}\footnote{\href{https://huggingface.co/sentence-transformers}{https://huggingface.co/sentence-transformers}} is widely regarded as a state-of-the-art technique for generating high-quality sentence and text embeddings. Known for its ability to capture semantic similarities between documents, it is one of the most popular choices for this task. \emph{SentenceTransformers} offers numerous pre-trained models, all hosted on the Huggingface Model Hub \cite{huggingfacemodels}. Among these, the \emph{all-*} models were trained on a vast dataset of over one billion training pairs, making them highly versatile and suitable for a wide range of general-purpose applications. Additionally, there are \emph{Multi-QA} models, specifically trained on 215 million question-answer pairs from diverse sources, including StackExchange, Yahoo Answers, and search queries from platforms like Google and Bing. Notable models in this category include \emph{multi-qa-MiniLM-L6-dot-v1}, \emph{multi-qa-distilbert-dot-v1}, and \emph{multi-qa-mpnet-base-dot-v1}. The \emph{multi-qa-mpnet-base-dot-v1} model is renowned for its exceptional performance in semantic search tasks, while the \emph{multi-qa-MiniLM-L6-dot-v1} model is optimized for speed, offering a good balance between accuracy and efficiency.

For our embedding needs, we select the \emph{multi-qa-MiniLM-L6-dot-v1} model. This choice was driven by the model's training on question-answer data, which closely aligns with the nature of our dataset, making it well-suited for capturing the nuances of the content we are analyzing.

In BERTopic, the parameter \emph{nr\_topics} controls the number of topics by merging them after their initial creation, allowing for a fixed number of topics. However, it is generally recommended to control the number of topics using a clustering model for more accurate results. To achieve this, we utilize the \emph{HDBSCAN} clustering model \cite{mcinnes2017hdbscan}, which efficiently identifies topic clusters based on density. To enhance the default topic representation, we employ \emph{CountVectorizer}\footnote{\href{https://scikit-learn.org/stable/modules/generated/sklearn.feature\_extraction.text.CountVectorizer.html}{https://scikit-learn.org/stable/modules/generated/sklearn.feature\_extraction.text.CountVectorizer.html}}, which improves the quality of topics by ignoring infrequent words and extending the n-gram range. Following previous research works \cite{li2021understanding, abdellatif2020challenges}, we utilize the unigram and bigram models.

BERTopic includes a special \emph{-1} topic to group documents that do not fit well with the identified topics. This can occur when a document contains excessive noise, such as stopwords or irrelevant information, or does not strongly connect to the core themes identified by the model. The \emph{-1} topic acts as a 'catch-all' for these outlier documents. However, since we thoroughly preprocessed the data to remove such elements before applying BERTopic, our main topic begins with the label \emph{-1}. Finally, we identify nine distinct topics using the Stack Overflow data. Our findings are detailed in the Results \ref{case-study-results} Section.

% The script for generating these topics is included in the replication package, and 

\subsection{OpenAI Developer Forum}
\textbf{Step1: Crawl data from OpenAI developer forum. }We collect post data from the OpenAI Developer Forum, where each post typically includes key components such as a title, tags, category, question description, and often a code snippet. Additionally, each post is accompanied by metadata, including creation time, last reply time, view count, reply count, and the number of participating users. Using a Python script, we extract 26,474 unique posts from the forum. The platform's user base has been growing steadily, exceeding 800,000 users during our data collection period. This comprehensive dataset serves as the foundation for our analysis.
\newline
\textbf{Step 2: Preprocessing} To prepare the developer forum data for BERTopic modeling, we follow a structured preprocessing approach. First, we clean the text by removing irrelevant elements like punctuation, special symbols, and numbers that do not aid in topic understanding. We then normalize the text by converting it to lowercase for consistency. Using the NLTK stopwords list \cite{bird2006nltk}, we tokenize the text and eliminate common stopwords, as well as domain-specific terms like 'openai' and 'bill'. Additionally, we apply lemmatization to reduce words to their base forms. After this thorough cleaning, the preprocessed text is ready for topic generation using BERTopic.
\newline
\textbf{Step 3: Identify the topics: }We follow a similar approach to the one used for identifying topics in Stack Overflow posts \ref{so-topics} to uncover topics within the OpenAI Developer Forum data. By applying the BERTopic modeling technique, we successfully extract 17 distinct topics from the forum discussions. A comprehensive analysis of these topics, along with detailed results, is presented in the Results \ref{case-study-results} Section.
% \newline
% \textbf{The replication package of our study can be found using \cite{}}

\section{Results}\label{case-study-results}
In this section, we provide a comprehensive analysis of Stack Overflow posts and topics related to LLMs alongside an in-depth examination of topics derived from the OpenAI Developer Forum data. This dual analysis addresses our research questions and offers insights into the key themes and discussions surrounding LLMs and developer interactions on the forums. By exploring both datasets, we aim to highlight the most prominent topics and trends, shedding light on the core issues and conversations shaping the community.

\subsection{RQ1: What topics are Large Language Model developers are asking about?}
\textbf{Motivation: }The development of LLMs presents unique challenges that set it apart from traditional software engineering. LLM developers require specialized expertise in areas such as natural language processing (NLP), deep learning, and model optimization—skills that are often unnecessary in conventional software development. In addition, the large-scale training of these models demands proficiency in data preprocessing, hardware optimization, and resource management. Ethical considerations, such as model bias and responsible AI use, are more pressing in LLM development due to the significant societal implications of deploying such models. As a result, the challenges faced by LLM developers differ markedly from those encountered by traditional software developers.
This research analyzes data from Stack Overflow and the OpenAI Developer Forum to uncover the topics and issues shaping the LLM development community. By examining these discussions, we aim to identify areas of LLM development that are gaining momentum or proving difficult to address, offering insights into the evolving landscape of LLM technology. This understanding can help pinpoint gaps in existing documentation and tools, highlight emerging challenges, and guide future research. Furthermore, analyzing these developer discussions offers valuable feedback for improving LLM technologies, enhancing support systems, and refining developer resources, ultimately driving the advancement of LLM development.
\newline
\textbf{Approach: }In this study, we employ BERTopic to uncover and analyze the different topics discussed by developers on Stack Overflow and the OpenAI Developer Forum, as outlined in Section \ref{study-design}. The process of topic labeling involves three authors who serve as annotators. Each annotator independently reviews the top 20 keywords associated with each topic and manually inspects a random sample of at least 30 posts per topic. Based on this analysis, the authors assign provisional titles that best encapsulate the theme of the posts. To ensure consistency and accuracy, the annotators then convene to discuss and refine the labels, ultimately reaching a consensus on the final titles for each of the nine identified topics. We follow a similar methodology for labeling topics in the OpenAI Developer Forum dataset.

We also examine the most popular topics among developers. To assess topic popularity, we use two complementary metrics that have been widely adopted in previous research on Stack Overflow: average views and average score \cite{abdellatif2020challenges, ahmed2018concurrency, bagherzadeh2019going, bajaj2014mining, nadi2016jumping}. For the OpenAI Developer Forum, due to the lack of scoring mechanisms, we relied solely on average views to measure topic popularity.
\begin{enumerate}
    \item \textbf{The average number of views (avg. views) }a post receives from both registered and unregistered users serves as a key indicator of its popularity. High view counts reflect strong interest among LLM developers and provide a meaningful measure of community engagement by showing which topics resonate most with the developer community.
    \item \textbf{The average score (avg. score)}  of posts is another metric for gauging popularity. On Stack Overflow, users can upvote posts they find insightful or valuable and downvote those that don't meet content standards. These votes are combined into an overall score, reflecting the post's perceived value within the community and offering insight into its quality and relevance.
\end{enumerate}
\begin{table}[htbp]
    \centering
    \scriptsize
    \caption{The LLMs topics, keywords, and their popularity for Stack Overflow data.}
    \label{tab:llmssotopics}
    \begin{tabularx}{\textwidth}{r|p{3.5cm}|p{6.6cm}|r|r|r}
    % \begin{tabularx}{\textwidth}{r|X|X|r|r|r}
        \toprule
        SL. & Topic & Keywords & \# Posts& AvgView & AvgSc \\
        \midrule
        1  &LLM Ecosystem and Challenges & use, model, try, error, code, langchain, work, run, file, like, llm, question, want, follow, create, text, answer, python, openai, document. & 4344 & 1871.65 & 1.20 \\
        2  & API Usage & api, openai, use, error, code, response, try, azure, openai api, work, gpt, key, request, chatgpt, function, model, stream, create, api key, ai. & 1782 & 2015.57 & 0.96 \\
        3  & Environment Management  & gym, environment, openai gym, use, action, try, reward, agent, error, code, learning, openai, state, run, game, env, step, reinforcement, make, gym environment. & 606 & 2352.54 & 1.79 \\
        4  & LLM Training with Frameworks   & model,  bert,  huggingface,  use,  train,  transformer,  token,  dataset,  fine,  tokenizer,  layer,  try,  output,  input,  sentence,  tune,  fine tune,  code,  classification,  training.& 440 & 2703.31 & 2.15 \\
        5  & Programming Constructs and LLM Integration & vector,  document,  chromadb,  use,  chroma,  store,  embedding,  langchain,  database,  query,  collection,  search,  metadata,  datum,  create,  db,  embed,  vectorstore,  vector database,  try. & 394 & 1912.26 & 1.08 \\
        6  & Llama Indexing and GPU Utilization &llama,  gpu,  model,  use,  run,  error,  index,  try,  llama\_index,  llama index,  code,  load,  file,  llm,  import,  gb,  memory,  llamaindex,  cuda,  work. & 361 & 1440.87 & 0.87 \\
        7  & Audio Transcription and Speech Recognition Automation &  whisper,  audio,  file,  transcribe,  use,  audio file,  transcription,  try,  error,  enable,  speech,  node,  text,  openai whisper,  code,  use whisper,  openai,  python,  work,  library. & 258 & 1726.44 & 1.01 \\
        8  & Langchain Development and Error Handling & langchain,  error,  import,  try,  use,  code,  python,  file,  py,  run,  follow,  version,  appdata,  packages,  site packages,  work,  site,  line,  module,  package. & 212 & 2512.04 & 1.28 \\
        9  & Agents and Tool Interactions &  langchain,  tool,  use,  agent,  chain,  prompt,  answer,  output,  question,  llm,  user,  chat,  want,  history,  code,  context,  input,  work,  response,  try.   & 196 & 1682.94 & 0.89 \\
        \bottomrule
    \end{tabularx}
\end{table}
\textbf{Results Obtained Using Stack Overflow: }Following the procedure outlined in \textbf{Step 5} of Section \ref{so-topics}, we identify nine distinct topics within the Stack Overflow data. Table \ref{tab:llmssotopics} provides an overview of these topics, including their associated keywords, the number of posts related to each topic, and their popularity, as measured by views and scores from developers. As shown in the table, developer inquiries span a range of topics related to LLM development, with varying levels of engagement across topics. Below, we discuss these topics with examples.
\newline
\textbf{LLM Ecosystem and Challenges: }This is the most dominant topic, and it revolves around the growing landscape of LLMs and their application in practical, real-world tasks. Developers leverage LLMs to answer questions, process documents, and generate text. However, working with LLMs often entails running and debugging code in languages like Python, where developers face challenges ensuring that models run smoothly without errors. For instance, a developer working on an LLM project encountered difficulties while installing the package 'llama-cpp-python', as highlighted in the Stack Overflow post titled \emph{'Error while installing python package: llama-cpp-python'}. Other common issues include handling large files, integrating various tools, and refining models to improve their ability to process and respond to queries effectively. An example of this is the Stack Overflow post \emph{'How to upload files with the OpenAI API?'}, where a developer encountered issues when trying to append file information to the API. Additionally, developers often struggle with the complexities of LLM workflows, such as designing pipelines for document processing and generating coherent, contextually relevant answers. This is evident in the post \emph{'How to work with OpenAI's maximum context length of 2049 tokens?'}, where a developer faced limitations related to input size.

Despite the enormous potential of LLMs, developers must address the need for careful tuning and error handling to ensure optimal performance, particularly in tasks that involve diverse input types and real-time question-answering. As the LLM ecosystem continues to grow, it presents both technical challenges and significant opportunities for innovation.
\newline
\textbf{API Usage: }This topic focuses on the practical challenges of working with APIs, particularly those related to AI models like OpenAI's GPT. Developers frequently use the OpenAI API to build applications that generate responses or power AI-driven features, such as chatbots using ChatGPT. However, they often encounter errors while integrating these APIs, with common issues like API key misconfigurations or request failures. For example, in the Stack Overflow post \emph{'OpenAI API error: This is a chat model and not supported in the v1/completions endpoint'}, a developer encountered an error due to selecting the wrong engine. The developer used \emph{gpt-3.5-turbo} in a code that worked with the GPT-3 endpoint but needed to switch to \emph{text-davinci-003} for compatibility.

Similar challenges arise when using APIs in cloud environments like Azure, where authentication or response stream errors can occur. A related example is \emph{'Why am I getting 404 Resource Not Found for my newly Azure OpenAI deployment?'} The developer received a 404 error due to using a GET method instead of POST and providing an incorrect URL.

Successfully implementing APIs requires correctly structuring requests, handling responses efficiently, and managing tasks like streaming data or creating custom AI functionalities. Debugging code is a common part of the process, as seen in the case \emph{'openai.error.APIConnectionError: Error communicating with OpenAI'}, where the issue was tied to a macOS environment, and the developer had to install the emph{Certificates.command} utility. Understanding how to properly use API keys, make requests, and work with AI models is crucial for building powerful and reliable LLM applications.
\newline
\textbf{Environment Management: }This topic focuses on managing and interacting with environments. Developers often use tools like OpenAI Gym to simulate environments where agents interact with the environment by taking actions, receiving rewards, and learning from these experiences. Proper management of these environments is essential for creating intelligent agents capable of performing complex tasks, such as playing games or navigating simulations. However, developers often face challenges in setting up and using these environments effectively. For instance, in the Stack Overflow post titled \emph{Error in importing environment OpenAI Gym'}, a developer encountered issues when importing the \emph{Breakout} game environment, receiving an \emph{Unable to find game 'Breakout'} error. This issue was resolved by installing the necessary dependencies, such as \emph{gym[atari]} or \emph{gym[all]}.

Common problems discussed under this topic include environment setup errors, action space mismatches, and reward feedback misconfigurations, which are often encountered when integrating environment management into LLM-based projects. One such example is the post \emph{'How to render OpenAI gym in Google Colab?'}, where the developer needed assistance rendering the OpenAI Gym environment in Colab. The solution involved installing \emph{xvfb} and other dependencies, such as \emph{pyvirtualdisplay} and \emph{piglet}.
In addition to code and environment-specific issues, developers face platform and operating system challenges. For example, a post titled \emph{How to install mujoco-py on Windows?'} highlights the difficulties developers face when working across different systems.

Discussions around environment management serve as a valuable resource for solving issues related to code execution, environment state management, and agent behavior fine-tuning, all of which are critical for the smooth functioning of reinforcement learning processes in AI and LLM applications.
\newline
\textbf{LLM Training with Frameworks: }This topic centers on the practical aspects of training LLMs using popular frameworks like Hugging Face. Developers frequently work with models such as BERT, which utilizes transformer architectures, to tackle tasks like sentence classification and token generation. For instance, in the post \emph{'How to add new special token to the tokenizer?'}, a developer building a multi-class classification model struggled to add a new special token. The issue was resolved by using the \emph{add\_special\_tokens} method rather than \emph{add\_tokens}.

A typical workflow involves training or fine-tuning these models on datasets, using tokenizers to preprocess the input data into tokens for model comprehension. The training process includes configuring model layers, managing outputs, and fine-tuning to improve task-specific performance. An example of this is the post \emph{'How to get intermediate layers' output of a pre-trained BERT model in HuggingFace Transformers library?'}, where the developer needed to enable \emph{config.output\_hidden\_states=True} to access the intermediate layers.

Developers often experiment with input-output configurations and fine-tuning hyperparameters to optimize model performance. In the post \emph{'How to test a model before fine-tuning in Pytorch Lightning?'}, a developer explored methods to evaluate a model before the fine-tuning stage. Discussions frequently cover topics such as managing tokenization, adjusting transformer layers, and resolving issues related to classification and output. An example of such a discussion is \emph{'How padding in HuggingFace tokenizer works?'}.
\newline
\textbf{Programming Constructs and LLM Integration: }This topic focuses on integrating programming tools with LLMs, particularly for managing and querying data. A central focus is using vector databases like ChromaDB, which store document embeddings for efficient search and retrieval. For instance, in the post titled \emph{'LangChain Chroma - load data from Vector Database'}, a developer was able to store data but encountered difficulties loading it for future prompts. The solution was to define a retriever and pass it to a chain, a common practice in managing vector stores.

Developers often use tools like Langchain to build pipelines that connect LLMs with vector databases, allowing applications to embed, store, and query vast datasets. By embedding documents as vectors, systems can perform similarity searches across collections, improving the retrieval of relevant information based on metadata and vector representations. In the post \emph{'Dynamically add more embedding of a new document in Chroma DB - Langchain'}, a developer faced challenges in dynamically adding document embeddings. The issue was resolved by initializing a document object and appending a new list of document objects to the database. Typical programming constructs in this area include creating and managing databases, embedding data, and efficiently querying stored vectors for tasks like document search. For example, in a post titled \emph{'How ChromaDB querying system works?'} a developer sought to understand the querying process of ChromaDB. 

Developers frequently face challenges in setting up vector databases, organizing collections, and optimizing searches, making these topics common discussion points within the LLM development community. One such example is \emph{'Query existing Pinecone index without reloading the context data'}, where a developer sought advice on querying a vector database without reloading the context data.

\textbf{Llama Indexing and GPU Utilization: }This topic explores the challenges and techniques developers face when working with the Llama model, especially in GPU environments. Common issues include memory overloads and CUDA-related errors when handling large datasets or files. For instance, in a post titled as\emph{'llama-cpp-python not using NVIDIA GPU CUDA'}, a developer encountered difficulties utilizing the \emph{NVIDIA GPU CUDA}, which was resolved by installing the \emph{CUDA toolkit} and ensuring the correct path to the \emph{libllama.so} shared library. Exporting this library before running the Python interpreter addressed the problem effectively.

LlamaIndex (llama\_index) is critical for efficient data querying and indexing, but optimizing it to fully leverage GPU resources can be challenging. One example is \emph{'ModuleNotFoundError for llama\_index.vector\_stores'}, where the solution involved installing the missing \emph{llama-index-vector-stores-postgres} package. Similarly, in \emph{'How to properly import llama-index classes?'}, one developer sought help for proper imports within their project.

This topic also covers fine-tuning Llama models using Hugging Face Transformers. For example, in the post \emph{'Fine-tuning TheBloke/Llama-2-13B-chat-GPTQ model with Hugging Face Transformers library throws Exllama error'}, the issue was resolved by using the \emph{device\_map} attribute in the \emph{from\_pretrained} function. Through these discussions, developers explore essential techniques and troubleshoot common errors related to Llama model indexing and GPU optimization.

\textbf{Audio Transcription and Speech Recognition Automation: }This topic explores the challenges of using advanced tools like \emph{OpenAI Whisper} for speech-to-text conversion, audio transcription, and speech recognition automation. Developers frequently integrate Whisper to process audio files and efficiently transcribe speech across various applications. However, implementation challenges often arise, such as handling file formats or resolving errors during transcription. For example, one developer encountered the error \emph{'FileNotFoundError: [WinError 2] The system cannot find the file specified'} while using Whisper in Python, which was resolved by installing \emph{ffmpeg} as a Python package. Integration in non-Python environments adds complexity, as seen in posts like \emph{'How to use OpenAI Whisper in C\#?'}, where developers sought guidance on using Whisper in other programming languages.

Dependency and compatibility issues are also common. For instance, the error \emph{'FP16 is not supported on CPU; using FP32 instead'} was resolved by reinstalling \emph{pytorch} with CUDA and using the \emph{to} method to enhance performance. Library import problems, such as the post \emph{OpenAI Whisper Cannot Import Numpy'}, further highlight the importance of proper dependency management. These challenges reflect the nuanced efforts required to integrate and optimize Whisper across diverse development contexts effectively.

\textbf{Langchain Development and Error Handling: }This topic focuses on the challenges developers encounter when working with Langchain, a versatile framework for integrating language models into applications. Langchain is frequently used with Python to build applications that leverage LLMs for tasks like text generation, question answering, and natural language processing. Developers often run into issues related to importing necessary modules, handling execution errors, and resolving compatibility problems between different versions of Python or installed packages. A common issue, for example, is an import error like \emph{'langchain: No module named langchain.document\_loaders'}, which often arises from incorrect installation paths or missing packages.

Error handling is a vital part of Langchain development, as developers must ensure their applications run smoothly by troubleshooting problems related to environment setup, package dependencies, and module imports. Version conflicts in libraries are also frequent, with errors such as \emph{'ERROR: Could not find a version that satisfies the requirement pandas$>=1.2.3'$}, which can typically be resolved by adjusting package versions or reinstalling the necessary dependencies. Additionally, developers face challenges when running Langchain code in different environments, often needing to fine-tune their code or manage dependencies to avoid runtime issues. An example of this is \emph{'Exception: java.lang.NoClassDefFoundError: jakarta/servlet/Filter'}, which highlights the importance of addressing environment-specific errors to maintain application functionality.
\newline
\textbf{Agents and Tool Interactions:} This topic highlights the challenges developers face when integrating agents with various tools. A key issue is ensuring that agents effectively manage complex user prompts and generate accurate responses by interacting with tools or external APIs. For instance, in the post titled \emph{'I can't get the langchain agent module to actually execute my prompt'}, the developer struggled with prompt execution, with suggestions involving the proper use of \emph{PREFIX}, \emph{SUFFIX}, and \emph{FORMAT\_INSTRUCTION} to guide agent behavior.

Developers also face difficulties in configuring agents to handle dynamic inputs, especially when contextual understanding across multiple queries or historical interactions is required. A related challenge, seen in the post \emph{'langchain with context and memory'}, involves using conversation memory to retain past interactions. Here, the developer could use \emph{ConversationalRetrievalChain} to incorporate context and input document retrieval.

Further complications arise in troubleshooting how agents interact with chains of tools, leading to issues like incorrect outputs or unexpected tool behavior. In one post, \emph{'Langchain: Why does langchain agent return the action input instead of running it?'}, the developer faced problems with action execution. Another example, \emph{'Adding an argument to Tool function in Langchain agents'}, shows challenges in customizing tool functionality. Some developers also report agents not using tools correctly, as seen in the post \emph{'Langchain agent does not always use the tool correctly'}.

These challenges intensify when agents need to execute diverse tasks such as code execution, question answering, or multi-step reasoning, requiring developers to fine-tune interactions between agents, tools, and LLMs meticulously.

\begin{boxK}
LLM developers inquire about various aspects of LLM development, including API usage, environment management, training models, and error handling. Based on the number of posts, average views, and average score, the most popular topic is \emph{LLM Training with Frameworks}. Although it has fewer posts (440), it boasts the highest average views (2,703.31) and score (2.15), indicating that each post garners significant attention and positive feedback. In contrast, while \emph{LLM Ecosystem and Challenges} has the most posts, its average views (1,871.65) and score (1.20) are lower, suggesting that the community finds training frameworks more valuable. Another popular topic is \emph{Environment Management}, with 606 posts, a high average view count (2,352.54), and a score of 1.79. This suggests a strong interest in discussions about managing environments like \emph{OpenAI Gym}, with high engagement.
\end{boxK}
\begin{table}[ht]
\scriptsize
\centering
   \caption{The LLM topics, keywords, and their popularity for Open AI Developer Forum data}
   \label{tab:llmopenaitopics}
   \renewcommand{\arraystretch}{0.65}
   % \begin{adjustbox}{max width=\textwidth}
    \begin{tabular}{r|p{3.5cm}|p{7.5cm}|r|r}
	\toprule
	SL. & Topic & Keywords & \# Posts & AvgView \\
	\midrule
	1  & API Usage and Error Handling & api, error, use, assistant, text, code, response, model, file, json, thread, account, request, whisper, dall, playground, message, output, davinci, create & 10213 & 2102.27  \\
	2  & LLM Functionalities & gpt, turbo, custom, 4o, gpt4, api, model, use, vision, preview, gpt3, action, access, image, fine, mini, response, error, prompt & 4673  & 2425.06  \\
	3  & Managing API Requests and Responses & api, assistant, key, error, use, run, response, request, message, work, create, thread, access, issue, function, provide, user, return, make, tool & 2087  & 1788.03  \\
	4  & Integrating ChatGPT into Web and App Services & chatgpt, plugin, api, use, app, plus, chatgpt4, new, access, code, response, prompt, help, make, web, chatgpts, google, integrate, user, issue & 1477  & 3373.4  \\
	5  & AGI and AI Project Collaboration & question, help, project, community, answer, topic, problem, new, solve, post, thank, idea, need, ask, forum, article, information, story, agi, make & 1155  & 1640.25  \\
	6  & Chat Completion and Bot Development & chat, chatbot, completion, bot, history, api, conversation, chatcompletion, create, use, response, user, message, assistant, context, build, way, playground, role, make & 1071  & 2299.26  \\
	7  & Intelligent Agents and Generative Models & ai, open, intelligence, artificial, generative, use, create, project, development, agent, model, build, tool, need, human, future, assistant, want, help, look & 927  & 1880.64  \\
	8  & Token Usage & token, limit, rate, exceed, usage, max, tokens, quota, ratelimiterror, count, increase, error, current, reach, request, plan, check, context, maximum & 761  & 2436.67  \\
	9  & Fine-Tuning and Dataset Management & fine, tuning, tune, model, tuned, job, dataset, use, answer, training, loss, train, question, data, result, create, fail, babbage, example, work & 737  & 1389.61  \\
	10 & File Management and Retrieval & file, upload, assistant, search, api, retrieval, attach, uploaded, use, retrieve, delete, uploading, unable, pdf, error, attachment, vector, store, content, tool & 575  & 1416.93  \\
	11 & Prompt Engineering & prompt, engineering, design, help, fine, text, response, language, multiple, result, good, generate, create, generation, user, different, chain, tuning, single, need & 491  & 2378.94  \\
	12 & ChatGPT Models & chat, gpt, chatgpt, chatbot, use, turbo, api, custom, gpt4, plus, vs, gpt3, 4o, completion, version, website, response, model, different, access & 417  & 3708.47  \\
	13 & Plugin Development & plugin, access, develop, store, install, unverified, manifest, developer, development, install, plug, test, option, work, waitlist, browse, approve, long, available, submission & 410  & 1019.00  \\
	14 & Image Generation & image, generation, generate, edit, text, dall, api, dalle, url, prompt, base64, use, create, variation, input, description, photo, picture, png, error & 404  & 2016.11  \\
	15 & Embedding & embed, embedding, ada, text, similarity, model, vector, search, use, vs, large, question, small, semantic, create, input, dimension, new, util & 384  & 2545.79  \\
	16 & Data Preparation and Structured Analysis & datum, analysis, training, fine, structured, use, preparation, tune, api, tabular, tool, good, format, assistant, prepare, large, train, extract, cli, model & 347  & 1893.38  \\
	17 & Function Parameters and Callback Handling & function, parameter, argument, parallel, tool, multiple, require, return, response, use, type, value, callback, output, result, description, force, array, action, enum & 345  & 1789.2 \\ 
\bottomrule 
\end{tabular}
% \end{adjustbox}
\end{table}

\textbf{Results Obtained Using OpenAI Developer Forum: }We follow a similar process to extracting relevant topics from the OpenAI developer forum data, such as Stack Overflow data. Table \ref{tab:llmopenaitopics} provides an overview of these topics, highlighting their associated keywords, the number of posts for each topic, and their popularity based on the number of views from developers. As shown in Table \ref{tab:llmopenaitopics}, LLM developers face a wide range of challenges across various aspects of development, illustrating the diversity of issues in this field. By analyzing data from the OpenAI developer forum, we identified 17 distinct topics. Below, we provide an overview of each category along with examples, followed by a summary of our findings.

\textbf{API Usage and Error Handling: }This topic focuses on the challenges developers face when integrating and interacting with APIs, particularly in the context of AI models like OpenAI's assistant and tools such as Whisper and DALL-E. Common issues include malformed requests, unexpected API responses, and difficulties handling JSON data or file operations. Developers often need to troubleshoot response codes, manage authentication (e.g., API keys), and resolve errors related to model outputs. For instance, one developer encountered an SSL error \emph{'SSL: certificate\_verify\_failed'} while connecting to the OpenAI API from Python, which required manually adding certificates to the cacert.pem file.

The topic also includes API integration into development environments like playgrounds, where developers manage model features such as output generation, file handling, and message threading. An example from the forum involved a question titled \emph{'How do I use the new JSON mode?'}. Error handling is particularly crucial when working with models like Davinci or Whisper, as smooth interactions between components are necessary for effective AI usage. For instance, a developer faced the \emph{'Error retrieving completions: 400 Bad Request'} error due to an incorrect \emph{max\_tokens} parameter, which was resolved by adjusting the value.

\textbf{LLM Functionalities: }This topic emphasizes the practical use and customization of GPT models (e.g., GPT-3, GPT-4) across diverse applications. Developers often interact with APIs to tailor model behavior, manage actions through prompts, and perform vision-based tasks like image processing. For example, a query titled \emph{'How to make an API call to a custom GPT model?'} received guidance on leveraging the Assistant API for customization.

This topic also discusses the challenges of fine-tuning and integrating models like GPT Turbo to enhance performance. Common issues include managing API responses, troubleshooting errors, and accessing models, as seen in posts like \emph{'How to deal with 'lazy' GPT-4?'}. Developers are exploring advanced features, such as vision capabilities and preview versions, seamlessly combining text and image processing. One notable post, \emph{'GPT-4 API and image input'}, highlighted the demand for documented methods to supply images to the GPT-4 API, showcasing the evolving applications and needs in GPT model development.

\textbf{Managing API Requests and Responses: }This topic focuses on the challenges and best practices for handling API interactions effectively. This includes managing API keys, troubleshooting errors, creating and executing requests, and accurately interpreting responses. For instance, in the post titled \emph{'Issue with Accessing choices Attribute from OpenAI API Response'}, a developer encountered an error mentioning \emph{member choice is unknown'}. The issue was due to a typo where \emph{choice} was used instead of the correct attribute \emph{choices}.

Developers often face challenges like unexpected behavior of APIs, incorrect responses, access issues, and system errors, which can impact application functionality. An example of this is \emph{'API aborts my connection without a reason - anything I can do?'}, highlighting the struggle with unexplained connection issues. Additionally, ensuring reliable and efficient communication between client and server is a key aspect of managing APIs. One common problem involves getting the API to follow instructions accurately, as illustrated in \emph{'How to force the API to really follow all the instructions?'}, where a developer experienced inconsistent behavior from the LLM API. This topic aims to provide developers with insights and best practices for creating robust and resilient API-based systems, addressing these common challenges comprehensively.

\textbf{Integrating ChatGPT into Web and App Services: }This topic focuses on developers' various challenges when incorporating ChatGPT functionalities into their applications. One major issue is managing API limitations and ensuring seamless plugin integration, as illustrated by a developer's question: \emph{'Do plugins work via API or only ChatGPT?'}. This query highlights the need for clarity around accessing plugins via API calls. Similarly, developers discussed using ChatGPT in different web services, such as the question \emph{'ChatGPT/OpenAI for PHP website'}, where a developer inquired about building OpenAI functionality into a PHP-based platform. Another example we found in this topic is \emph{How do I call ChatGPT API with Python code?},  where one developer is searching the code to call the ChatGPT API for the developer application. Other challenges include prompt handling, customizing ChatGPT to meet specific app requirements, and resolving compatibility issues with other platforms, such as Google services. An example of this is the question \emph{'How to show images on ChatGPT plugins?'}, which points to the difficulties in integrating media handling within the ChatGPT environment.

Furthermore, complexities arise in managing user authentication, access control, and balancing API usage limits with other app services. A notable example is the query \emph{'How can I link to ChatGPT web prepopulating the first prompt on the chat?'}, which reflects the technical hurdles of creating smooth user experiences in real-time interactions. These examples highlight the multifaceted challenges developers must navigate when integrating ChatGPT into web and app services.enges developers must navigate when integrating ChatGPT into web and app services.

\textbf{AGI and AI Project Collaboration: }This topic focuses on the challenges developers face while working on Artificial General Intelligence (AGI) and broader AI projects, particularly in the context of community-driven problem-solving. Developers often turn to forums like the OpenAI developer community to seek help, ask questions, and share solutions for complex issues. These discussions range from tackling specific technical problems to sharing ideas, articles, and stories that offer insights into new approaches. One such example is \emph{'Finetuning for Domain Knowledge and Questions'}, where one developer provided an excellent discussion for finetuning.

 The collaborative nature of the forum fosters an environment where community members can solve problems collectively, offering guidance on project-related challenges, and contributing to the development of AGI and AI innovations. One such post from a developer was \emph{'VERSES Declares Path to AGI, Now What?'}. This topic highlights the importance of crowdsourced knowledge and the role of developer communities in overcoming obstacles and advancing AGI projects.

 \textbf{Chat Completion and Bot Development: }This topic focuses on the challenges developers encounter while building and managing chatbot systems that rely on API-driven chat completion features. For instance, one issue raised on the OpenAI developer forum is titled \emph{'AttributeError: module openai has no attribute ChatCompletion'}, highlighting configuring API issues. A significant challenge involves creating bots that maintain coherent conversations by effectively utilizing context and managing user interaction history. One example is a post titled \emph{'How to provide context in a Q\&A chatbot?'}, where developers discuss difficulties in sustaining relevant conversations over time. Another issue arises from navigating API limitations to ensure accurate, timely, and contextually appropriate responses. For example, one developer asked for help with \emph{'Configuring timeout for ChatCompletion Python'}, seeking guidance on how to set appropriate timeouts.

Balancing bot roles, understanding user intent, and managing complex conversational flows are further hurdles. One such challenge, \emph{'OpenAI API Conversation Memory'}, involves handling conversation continuity across multiple interactions. Additionally, developers face difficulties integrating these systems into platforms like OpenAI Playground and ensuring smooth functionality in real-world applications, such as troubleshooting inconsistent response times, as seen in the post \emph{'Slow Chat API responses'}. These examples illustrate the complexity of developing bots that deliver human-like conversations, requiring developers to balance API usage, response quality, and system performance.

\textbf{Intelligent Agents and Generative Models: }This topic explores AI-driven systems and generative technologies, emphasizing the development of agents capable of complex tasks and adapting to dynamic environments. One of the primary challenges is designing models that generate relevant and accurate outputs while maintaining the flexibility to adapt to diverse use cases. For example, a developer in the forum raised the question, \emph{'What are some strategies to bypass GPTZero or other AI detection tools?'}, reflecting the complexity of generating outputs that can bypass detection mechanisms. 

Developers also faced issues related to scalability and robustness, ensuring that these systems can manage real-time interactions, anticipate human requirements, and make informed decisions. For instance, a post titled \emph{'Are You Intelligent Enough To Become An AI Enhanced Human?'} touches on the futuristic considerations of AI augmentation, while another forum post, \emph{'How to extract technical expressions from PDFs so that they can be understood by AI?'}, underscores the technical challenges developers face in making AI systems more effective.

\textbf{Token Usage: }This topic focuses on developers' challenges in managing token limits and quotas in API-driven systems. Developers frequently encounter issues such as exceeding token limits, dealing with rate-limiting errors, and balancing token usage across multiple requests. For example, one developer raised the question \emph{'Context length vs Max token vs Maximum length'}, seeking clarification on the differences between these three concepts.

A common challenge is ensuring API calls remain within the maximum token limit while optimizing resource use. Additionally, developers must track token consumption and address errors like \emph{ratelimiterror} when thresholds are exceeded. One such query, \emph{'What is the OpenAI algorithm to calculate tokens?} exemplifies this struggle. This topic underscores the need for effective strategies to monitor, manage, and optimize token usage while avoiding disruptions in API functionality.

\textbf{Fine-Tuning and Dataset Management:} This topic explores the challenges developers encounter when fine-tuning models and managing the datasets required for this process. Optimizing the tuning process to achieve the desired model performance while minimizing training loss and errors is a common difficulty. For example, one developer raised the question \emph{'Steps Meaning in Fine-Tuning and How to Pick the Optimal Number of Epochs from Them?'}.

Developers often face issues such as training job failures, inadequate datasets, and selecting the appropriate examples for fine-tuning. Understanding the intricacies of how fine-tuning works is another challenge, as reflected in the query \emph{'How does fine-tuning really work?'}. Moreover, handling large datasets and overcoming problems like failed tuning jobs or unsatisfactory results add further complexity. One developer, for instance, asked about \emph{'Preparing data to fine-tune a function-calling model'}. This topic underscores the importance of effective dataset management strategies for fine-tuning and model optimization.

\textbf{File Management and Retrieval: }This topic examines the challenges developers encounter when managing files in LLM systems. Common issues include file uploads, retrievals, and handling errors like upload failures or retrieval disruptions. For instance, one developer reported being \emph{'Unable to upload documents'} while attempting to upload a PDF to a custom GPT model. Others faced retrieval challenges, as highlighted in posts like \emph{'Assistant API, retrieval file API is not working'}.

The complexity increases when integrating file management with advanced functionalities such as vector storage and content search for smooth access and retrieval. For example, a developer raised concerns about implementing \emph{File search + function calling on Assistants'}, emphasizing the challenge of seamless integration. This topic highlights the need for robust strategies and tools to ensure efficient file management and address API-based file-handling issues effectively.

\textbf{Prompt Engineering: }This topic discusses the challenges developers face in designing and refining prompts for language models. Effective prompts are essential for generating precise and high-quality responses, leading developers to experiment with various designs to achieve optimal results. For example, a developer asked \emph{'Best prompt for generating precise TEXT on DALL-E 3'} to identify the ideal prompt structure. Another developer explored ways to enhance response accuracy with \emph{'A better Chain Of Thought prompt'}. Challenges also include tailoring prompts to produce coherent and user-specific responses, as illustrated by the question \emph{'Contextualizing completions: fine-tuning vs. dynamic prompt engineering using embeddings'}. Moreover, developers need strategies for chaining prompts to support complex tasks and workflows, ensuring response consistency and clarity. This topic emphasizes the pivotal role of well-crafted prompts in improving the performance and usability of language models.

\textbf{ChatGPT Models: }This topic highlights various challenges developers face when working with different ChatGPT versions. Choosing the appropriate model—whether GPT-3, GPT-4, or the GPT-4 Turbo variant—presents unique considerations, as each version offers distinct response styles, processing speeds, and pricing structures. For instance, one developer raised a question titled \emph{'GPT-3.5-turbo: How to Remember Previous Messages Like Chat-GPT Website'}, noting that while the official ChatGPT website retains conversation context, this feature isn't the same in the API usage. Similarly, developers are eager to extend ChatGPT functionalities to other applications, like Google Docs, as reflected in the post \emph{'Chat GPT for Google Docs, Gmail, Sheets, and Slides'}. Additionally, differences in model quality also raised discussion, with one developer asking \emph{'ChatGPT vs GPT-4o via APIs: Noticeable Quality Difference'}, observing that ChatGPT's web interface yields better responses than the API and seeking clarity on why. As they navigate these hurdles, developers strive to enhance model performance, optimize token usage, and deploy scalable, cost-effective applications across platforms, underscoring the demand for adaptable solutions within the ChatGPT ecosystem.

\textbf{Plugin Development: }This topic explores various challenges developers face in creating, submitting, and managing plugins. Common issues include navigating plugin access permissions, approval processes, and platform-specific integration requirements. For example, developers often encounter errors when dealing with unverified plugins, as seen in the post \emph{'Message: "ERROR UNINSTALLING (plugin name)" with unverified plugins – is this malware?'}. Installing and testing plugins frequently presents its own complexities, with errors like \emph{ChefBuildError} causing installation failures, as described in \emph{'Installing the retrieval plugin'}. Additionally, some developers struggle to set up plugin development in local environments, prompting posts like \emph{'Can't develop plugin locally'}, with requests for guidance on enabling development capabilities. This topic captures the evolving needs and hurdles in plugin development, offering insights into optimizing the entire plugin lifecycle—from initial creation to user adoption.
\begin{boxK}
     Table \ref{tab:llmopenaitopics} presents 17 topics identified from the OpenAI Developer Forum. \emph{API Usage and Error Handling} emerges as the most discussed topic with 10,213 posts, indicating its critical relevance among developers, while \emph{ChatGPT Models} has the highest average views (3,708.47), reflecting substantial interest in understanding different versions and their functionalities. Topics such as \emph{LLM Functionalities} and \emph{Integrating ChatGPT into Web and App Services} also rank highly, focusing on optimizing and implementing ChatGPT capabilities in various applications. Other notable areas include \emph{Token Usage} and \emph{Prompt Engineering}, which receive moderate attention but highlight specific technical challenges developers face. Overall, the results strongly emphasize API management, model customization, and integration challenges in the developer community.
\end{boxK}
\textbf{Image Generation: }This topic investigates various challenges developers face when working with text-to-image generation, particularly with the DALL-E API. A recurring issue is unintended prompt alterations, as illustrated by the post \emph{'API Image Generation in DALL-E-3 changes my original prompt without my permission'}. Key difficulties also involve composing effective prompts, managing API calls, and handling image variations, formats, and errors. For example, one developer inquired about achieving consistent style with modified poses, posting \emph{'Prompt to make exactly the same image but different pose'}. Troubleshooting output issues, such as orientation for vertical images, is also a frequent topic, with questions like \emph{'Orientation problem for vertical images'}. Furthermore, developers often seek advice on image editing and format compatibility, particularly for PNGs, as seen in the post \emph{'How can I provide an RGBA PNG file to OpenAI PHP library?'}. This topic discussed the complexities of achieving high-quality visual outputs, highlighting common hurdles in image generation and editing workflows.

\textbf{Embedding: }This topic explores the challenges of creating and utilizing text embeddings to capture semantic meaning. It focuses on models like OpenAI's Ada-002, which converts text into dense vector representations for similarity search, semantic matching, and clustering tasks. Developers often face difficulties understanding and implementing embeddings, as highlighted by posts like \emph{'Understanding text-embedding-ada-002 vector length of 1536'} and \emph{'How to deal with different vector dimensions for embeddings and search with pgvector?'}. Performance and efficiency issues also frequently arise, as seen in queries like \emph{'Semantic embedding: super slow "text-embedding-ada-002"'}. Understanding how to create, apply, and fine-tune embeddings effectively empowers developers to build more intelligent, responsive systems that leverage the nuances of natural language data.

\textbf{Data Preparation and Structured Analysis: }This topic involves organizing, cleaning, and transforming raw data into structured formats suitable for analysis and model training. For instance, a developer raised a query titled \emph{'Getting response data as a fixed \& consistent JSON response'}, where they sought a standardized JSON format for further analysis. This topic covers various tasks, such as data extraction from diverse sources using APIs and CLI tools and converting tabular data into standardized formats for applications like embedding. An example is a developer post titled \emph{'Preparing data for embedding'}. Developers also need to focus on fitting data to specific models, including steps like scaling, encoding, and handling missing values, as seen in the post \emph{'Summarizing and extracting structured data from long text'}. Challenges with data extraction from various sources are also common, with developers posting about issues like \emph{'Extracting Data From PDFs'} and \emph{'Tabular data converted to embeddings not returning accurate results'}.

\begin{boxK}
\textbf{RQ1 Summary:} We identify the LLM developer challenges using OpenAI Developer Forum and Stack Overflow. The \emph{API Usage} topic emerges as a common and significant discussion area on both platforms, indicating persistent challenges in managing and implementing APIs for LLM-based applications. On the OpenAI Developer Forum, the most prominent topic is \emph{API Usage and Error Handling} with the highest post count (10,213), showing the community's engagement with troubleshooting and optimizing API performance. On Stack Overflow, \emph{LLM Ecosystem and Challenges} has the most posts (4,344), reflecting broader challenges developers encounter while integrating and troubleshooting LLMs in various scenarios.

While the OpenAI forum has high average views for \emph{ChatGPT Models} (3,708.47), demonstrating an intense focus on understanding and deploying specific model versions, Stack Overflow's \emph{LLM Training with Frameworks} has the highest average views (2,703.31), suggesting a strong interest in model training methods, particularly with frameworks like Hugging Face. Other topics such as \emph{Environment Management}, \emph{Langchain Development}, and \emph{Prompt Engineering}, emphasize developers' focus on configuring environments, building effective prompts, and managing LLM agents and tool interactions. Our analysis reveals that API usage, model integration, and environment setup are the most prevalent challenges developers face when working with LLMs, with communities actively seeking solutions and sharing knowledge around these challenging areas.
\end{boxK}

\textbf{Function Calling: }This topic explores developers' challenges when implementing and managing function calls, focusing on handling diverse parameters, return types, and callback mechanisms. For example, a developer raised a query titled \emph{'Function Calling parameter types'}, struggling with defining parameter types while using an OpenAI function. Similarly, another post, \emph{'How to specify arguments for function calling interactively'}, highlights the need for clarity in structuring function signatures and aligning arguments with expected outputs.

This topic also includes difficulties with asynchronous calls, managing parallel processes, and using callbacks to handle multiple responses effectively, as seen in posts like \emph{'Callbacks for long-running processes?'}. Developers also encounter challenges in managing arrays, enumerations, and configuring function calls to return the desired response structure. Posts such as \emph{'Function calling not returning the expected response structure'} and \emph{'Function Calling - Return Multiple Objects'} illustrate these struggles. Additionally, passing arrays or lists as function parameters remains an issue, as reflected in posts like \emph{'Function calling, passing a list of values'}.

\subsection{RQ2: What types of questions are  Large Language Model developers are asking?}

\textbf{Motivation: }Building on the topics identified in \textbf{RQ1}, a crucial question emerges: \emph{What types of questions are LLMs developers actually asking?}. Examining the nature of questions developers post on Stack Overflow and OpenAI Developer Forum about LLMs reveals not only the content but also the intent behind these interactions. For Stack Overflow data, we categorize questions by types such as 'How', 'Why', and 'What', shifting the focus from merely cataloging inquiries to uncovering why LLM developers rely on community knowledge over official documentation alone. Prior research \cite{rosen2016mobile, abdellatif2020challenges, treude2011programmers} suggests that categorizing questions in this way uncovers distinct problem-solving strategies and cognitive approaches among developers. By analyzing these question types across LLM topics, we can identify frequent issues and, more importantly, the underlying complexities developers face during LLM development. Similarly, within the OpenAI Developer Forum, we identify types like 'Model Development and Deployment', 'Evaluation and Optimization', and 'Troubleshooting'. This detailed understanding aids in refining model documentation, user support, and educational resources, empowering OpenAI and other stakeholders to build features and tools that better meet developers' practical needs, streamline troubleshooting, and encourage broader, more effective LLM adoption.

\textbf{Approach: }To address this research question, we employ an approach consistent with prior methodologies for categorizing the types of Stack Overflow posts \cite{rosen2016mobile, abdellatif2020challenges, treude2011programmers}. Specifically, we select a statistically significant random sample of posts from each of the nine LLM-related topics, ensuring a 95\% confidence level and a 5\% confidence interval. This sampling yields a total of 1,918 posts distributed across topics as follows: 354 posts on LLM Ecosystem and Challenges, 317 on API Usage, 236 on Environment Management, 206 on LLM Training with Frameworks, 195 on Programming Constructs and LLM Integration, 187 on Llama Indexing and GPU Utilization, 155 on Audio Transcription and Speech Recognition Automation, 137 on Langchain Development and Error Handling, and 131 on Agents and Tool Interactions.

\begin{itemize}
    \item \textbf{How: }This category includes posts where developers request guidance on specific techniques or steps to accomplish a task, emphasizing practical, goal-oriented solutions \cite{rosen2016mobile, abdellatif2020challenges}. These posts typically seek clear, sequential instructions \emph{e.g., 'How to solve API connection error and SSL certification error while connecting to GPT-3 open AI?'}.
     \item \textbf{Why: }Posts in this category are centered on understanding reasons or explanations for certain phenomena \cite{rosen2016mobile, abdellatif2020challenges}. Often associated with troubleshooting, these questions focus on exploring the underlying causes of specific behaviors or issues \emph{e.g., 'Why can't I access GPT-4 models via API, although GPT-3.5 models work?'}
    \item  \textbf{What: }This category comprises posts where developers seek information to clarify doubts or make more informed decisions \cite{rosen2016mobile, abdellatif2020challenges}. These inquiries often aim at understanding best practices or identifying appropriate tools \emph{e.g., 'What does langchain CharacterTextSplitter's chunk\_size param even do?'}.
    \item \textbf{Others: }This category captures posts that don't fit into the above types, often involving requests for conceptual understanding or exploratory information  \cite{rosen2016mobile, abdellatif2020challenges} \emph{e.g., 'Understand Adam optimizer intuitively'}.
\end{itemize}

Two independent authors reviewed the title and body of each post to categorize its type, following a single-label assignment approach that prioritized the most relevant label when posts fit multiple categories. Discrepancies were resolved through collaborative discussion to reach a consensus, reinforcing the classification reliability. Additionally, we cross-checked our categorization criteria with the labeled dataset from \cite{abdellatif2020challenges}, which provided a robust framework for similar analyses in chatbot development, enhancing our consistency.

To evaluate the quality of our classification, we employ Cohen’s Kappa \cite{mchugh2012interrater}, a statistical measure of inter-rater reliability. This metric objectively assesses the agreement level among annotators, ensuring the consistency and robustness of the categorization process. Our result yields a kappa score of 0.87, indicating a notably high level of agreement among annotators. This strong reliability underscores the robustness and coherence of the annotated dataset.

We intend to analyze the OpenAI Developer Forum to uncover the specific types of questions LLM developers are asking. While Stack Overflow provides a comprehensive foundation for understanding developers' challenges, the OpenAI forum serves as a focused space that highlights nuanced or organization-specific inquiries. Although we initially consider applying the same classification schema used for Stack Overflow ('How', 'Why', 'What', and 'Other'), the nature of posts in the OpenAI forum reveals the need for a distinct set of categories to better capture the forum's unique discourse.

Since the OpenAI Developer Forum encourages discussion among developers actively working with OpenAI's APIs and models, it also provides visibility into challenges unique to this context, such as ethical considerations and feature requests that may be less apparent on Stack Overflow. To adapt to the forum's structure and context, we employ a mixed-method approach, combining natural language processing with expert review, to accurately capture the depth and variety of questions posed by the developer community on this platform.

After manually reviewing a significant sample of posts (450 posts) and leveraging ChatGPT-4o to brainstorm potential categories, we refine the categorization schema into eight distinct types:
\begin{itemize}
    \item \textbf{Model Development and Deployment}: Questions about technical tasks like fine-tuning, deployment, and data preprocessing.
    \item \textbf{Evaluation and Optimization}: Queries on performance benchmarking, model evaluation, and optimization strategies.
    \item \textbf{Ethical and Best Practices}: Posts discussing ethical concerns, standards, and best practices in model development.
    \item \textbf{Troubleshooting}: Real-time problem-solving inquiries that require immediate troubleshooting support.
    \item \textbf{Feature Requests and Improvements}: Suggestions for new features or enhancements in OpenAI models.
    \item \textbf{Guidance and Validation}: Requests for advice or validation on specific technical topics.
    \item \textbf{Maintenance and Ongoing Support}: Topics on sustaining or supporting existing models.
    \item \textbf{Other}: Posts that did not fit any of the above categories, capturing exploratory or general questions.
\end{itemize}
Using this schema, we employ OpenAI's GPT-4o model to classify each topic's posts. This approach involves crafting detailed prompts outlining each category, ensuring the model comprehensively understands each type. Our analysis begins by selecting a representative sample, drawn with 95\% confidence and a 5\% margin of error, from the 17 topics. The sampled posts are then categorized according to a refined schema, followed by a manual validation process to ensure accuracy. Through the expert review, we aim to achieve high classification reliability—which is essential given the nuanced nature of developer questions on this specialized platform.
\begin{table}[htbp]
    \centering
    \scriptsize
    \caption{LLM-related post types on Stack Overflow.}
    \label{tab:swssotopicstypes}
    \begin{tabular}{p{4.5cm} |r |r |r |r}
        \toprule
        \textbf{Topic} & \textbf{\% How} & \textbf{\% Why} & \textbf{\% What} & \textbf{\% Other} \\
        \midrule
        LLM Ecosystem and Challenges  & 79.04 & 7.93 & 12.75 & 0.28 \\
        API Usage  & 47.95 & 29.34 & 21.77 & 0.95 \\
        Environment Management  & 50.00 & 24.15 & 20.76 & 5.08 \\
        LLM Training with Frameworks  & 56.80 & 17.48 & 23.79 & 1.94 \\
        Programming Constructs and LLM Integration & 74.87 & 12.82 & 11.80 & 0.51 \\
        Llama Indexing and GPU Utilization & 48.92 & 39.79 & 9.14 & 2.15 \\
        Audio Transcription and Speech Recognition Automation & 60.42 & 20.49 & 17.71 & 1.38 \\
        Langchain Development and Error Handling & 68.29 & 22.07 & 11.03 & 2.60 \\
        Agents and Tool Interactions & 53.28 & 24.81 & 18.98 & 2.92 \\
        \midrule
        \textbf{LLM (all)} & 59.95 & 22.10 & 16.41 & 1.98 \\
        \bottomrule
    \end{tabular}
\end{table}

\begin{figure}[htbp]
\centering
  \vspace{-0.5em}
  \includegraphics[width=8cm]{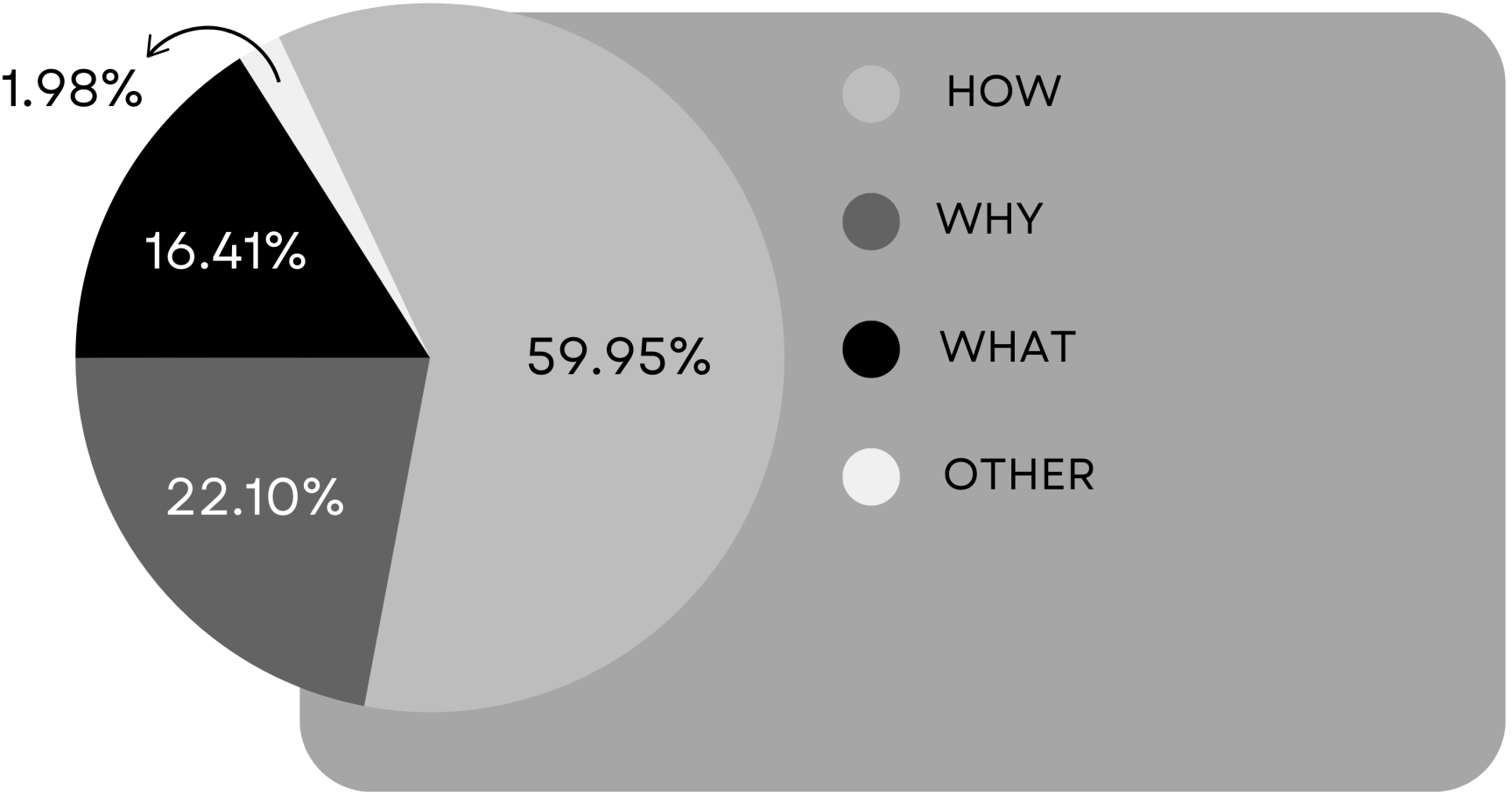}
  \caption{Types Distribution of SO Posts}
  \label{fig:typedistribution}
  \vspace{-0.8em}
\end{figure}

\textbf{Results: }Table \ref{tab:swssotopicstypes} shows the distribution of question types related to LLM topics on Stack Overflow. The categories analyzed are 'How', 'Why', 'What', and 'Other'. Overall, the majority of questions fall under the 'How' category, averaging 59.95\%, indicating that most inquiries revolve around practical implementation and procedural guidance. This is particularly evident in topics like \emph{LLM Ecosystem and Challenges} (79.04\%) and \emph{Programming Constructs and LLM Integration} (74.87\%). The 'Why' category, accounting for 22.10\% on average, reflects questions seeking explanations or underlying reasons, with a notable emphasis on \emph{Llama Indexing and GPU Utilization} (39.79\%) and \emph{API Usage} (29.34\%). The 'What' type of questions, averaging 16.41\%, showcases an interest in definitions and clarifications, with \emph{LLM Training with Frameworks} (23.79\%) and \emph{API Usage} (21.77\%) having higher proportions. The 'Other' category, encompassing 1.98\% of total posts, represents a minimal portion, although \emph{Environment Management} (5.08\%) and \emph{Agents and Tool Interactions} (2.92\%) show slightly higher figures. Overall, the data highlights a strong preference for practical 'How' questions, with varying levels of interest in theoretical 'Why' and explanatory 'What' types, while 'Other' remains consistently low across all topics. Figure \ref{fig:typedistribution} shows the summary of types distribution for our analysis.
We observe that 'How' questions consistently dominate across various domains, demonstrating a strong preference for practical and procedural inquiries. For example, 'How' questions make up 61.8\% of chatbot-related posts \cite{abdellatif2020challenges} and 59\% of mobile development posts \cite{rosen2016mobile}, closely aligning with the 59.95\% observed in our domain. The second most common type in chatbot and mobile development topics is 'Why' (25.4\% and 29\%, respectively), mirroring our findings, where 'Why' questions account for 22.10\% of the data. The 'What' category follows as the third most common type. Our analysis aligns with findings from similar studies conducted in other domains, reinforcing the consistency of these trends across different technical fields.
\begin{table}[htbp]
\centering
\scriptsize
\caption{Breakdown of Developer Challenge Types (Abbreviations: MDD - Model Development and Deployment, EO - Evaluation and Optimization, EBP - Ethical and Best Practices, TS - Troubleshooting, FRI - Feature Requests and Improvements, GV - Guidance and Validation, MOS - Maintenance and Ongoing Support, OTH - Other)}
\label{tab:developer_challenges-openaidevforum}
\begin{adjustbox}{max width=\linewidth}
\begin{tabular}{>{\raggedright\arraybackslash}p{5cm}|r|r|r|r|r|r|r|r}
\toprule
\textbf{Topic} & \textbf{\% MDD} & \textbf{\% EO} & \textbf{\% EBP} & \textbf{\% TS} & \textbf{\% FRI} & \textbf{\% GV} & \textbf{\% MOS} & \textbf{\% OTH} \\
\midrule
API Usage and Error Handling & 15.9 & 5.66 & 3.50 & 40.4 & 22.9 & 8.89 & 0.81 & 1.89 \\ 
LLM Functionalities & 29.7 & 4.21 & 3.65 & 28.4 & 19.1 & 12.9 & 0.56 & 1.40 \\ 
Managing API Requests and Responses & 19.3 & 0.38 & 1.85 & 49.2 & 20.9 & 6.76 & 0.92 & 0.62 \\ 
Integrating ChatGPT into Web and App Services & 23.2 & 5.29 & 2.92 & 13.4 & 37.9 & 13.7 & 0.98 & 1.63 \\ 
AGI and AI Project Collaboration & 9.34 & 5.29 & 8.99 & 25.6 & 20.0 & 22.5 & 1.73 & 8.30 \\ 
Chat Completion and Bot Development & 22.9 & 1.77 & 2.83 & 32.1 & 29.3 & 7.77 & 0.70 & 2.47 \\ 
Intelligent Agents and Generative Models & 17.2 & 2.20 & 11.1 & 19.8 & 25.3 & 16.1 & 0.73 & 7.35 \\ 
Token Usage & 12.5 & 6.64 & 0.78 & 67.9 & 9.37 & 1.95 & 0.39 & 0.39 \\ 
Fine-Tuning and Dataset Management & 72.7 & 4.35 & 0.79 & 17.8 & 1.98 & 1.58 & 0.39 & 0.39 \\ 
File Management and Retrieval & 12.9 & 1.29 & 1.29 & 62.3 & 19.9 & 1.30 & 1.30 & 0.43 \\ 
Prompt Engineering & 21.3 & 8.33 & 4.16 & 26.3 & 22.2 & 15.2 & 0.93 & 1.39 \\ 
ChatGPT Models & 32.9 & 3.34 & 2.86 & 29.5 & 22.9 & 6.67 & 0.47 & 1.42 \\ 
Plugin Development & 9.55 & 1.01 & 1.50 & 51.7 & 29.1 & 4.02 & 2.01 & 1.01 \\ 
Image Generation & 7.57 & 1.01 & 4.54 & 44.9 & 32.3 & 6.06 & 1.51 & 2.02 \\ 
Embedding & 26.4 & 17.6 & 1.55 & 22.2 & 21.8 & 8.29 & 0.51 & 1.55 \\ 
Data Preparation and Structured Analysis & 41.5 & 6.01 & 7.65 & 21.3 & 15.8 & 6.02 & 1.09 & 1.09 \\ 
Function Parameters and Callback Handling & 9.29 & 2.18 & 1.09 & 50.8 & 27.3 & 7.10 & 1.09 & 1.09 \\ 
\midrule
\textbf{Average} & 22.60 & 4.50 & 3.59 & 35.51 & 22.24 & 8.64 & 0.95 & 2.03 \\ 
\bottomrule
\end{tabular}
\end{adjustbox}
\end{table}

Table \ref{tab:developer_challenges-openaidevforum} provides a detailed classification of the OpenAI Developer Forum posts based on the eight refined categories. The results show distinct patterns in the types of questions developers ask. \emph{Troubleshooting} is the most frequent category, accounting for the majority of posts in topics such as Token Usage (67.9\%), File Management and Retrieval (62.3\%), and Plugin Development (51.7\%). This trend highlights the technical obstacles developers encounter when using OpenAI’s tools and APIs, emphasizing the community's need for immediate problem-solving resources.
\newline
\emph{Model Development and Deployment} is another prevalent category, particularly in Fine-Tuning and Dataset Management (72.7\%) and Data Preparation and Structured Analysis (41.5\%), underscoring the importance of foundational tasks in LLM workflows. These findings align with the Stack Overflow dataset’s dominance of ‘How’ questions, emphasizing a need for practical guidance in implementing LLM-related tasks. \emph{Feature Requests and Improvements} constitute a significant portion of posts in categories like Integrating ChatGPT into Web and App Services (37.9\%) and Image Generation (32.3\%). This indicates developers’ proactive involvement in suggesting enhancements and expanding capabilities.
\newline
\emph{Guidance and Validation}, though less frequent, remains relevant in topics such as AGI and AI Project Collaboration (22.5\%) and Prompt Engineering (15.2\%). These findings mirror the ‘Why’ questions from Stack Overflow, reflecting developers’ desire for validation and deeper insights into best practices. While categories like \emph{Ethical and Best Practices}, \emph{Evaluation and Optimization}, and \emph{Maintenance and Ongoing Support} are less represented, their presence highlights specialized discussions, particularly in Ethical Considerations (e.g., Prompt Engineering at 4.16\%) and Optimization Strategies (e.g., Embedding at 17.6\%).

Overall, these findings suggest that while developers are focused on overcoming immediate technical challenges, there is also significant engagement in enhancing current tools and validating their approaches. This insight can guide future efforts to address these areas through improved support mechanisms, targeted tutorials, and expanded documentation tailored to developer needs.

The results reveal a strong demand for practical guidance, troubleshooting resources, and innovative feature discussions across both datasets. The dominance of ‘How’ questions on Stack Overflow and Troubleshooting posts on the OpenAI Developer Forum highlights the universal need for immediate, actionable solutions. Similarly, the overlap between ‘Why’ questions and Guidance and Validation suggests developers’ shared interest in understanding and validating their workflows.
\newline
These insights emphasize the necessity for better support systems, comprehensive documentation, and community-driven enhancements to address the evolving challenges in LLM development. OpenAI-specific discussions further reveal developers’ proactive involvement in shaping tools and practices, offering valuable directions for improving support resources and fostering collaboration.
\begin{boxK}
\textbf{RQ2 Summary: }We have derived the following summary:
\newline
\textbf{Dominance of Practical Guidance}: Both datasets emphasize the need for practical, step-by-step solutions. Stack Overflow’s majority of ‘How’ questions (59.95\%) aligns with the high frequency of Model Development and Deployment and Troubleshooting posts in the OpenAI dataset. This indicates a universal demand for guidance across platforms, particularly in implementation and debugging tasks.
\newline
\textbf{Troubleshooting and Technical Challenges}: The prominence of troubleshooting questions in the OpenAI dataset (e.g., Token Usage at 67.9\% and File Management at 63.2\%) reflects the ‘Why’ questions observed in the Stack Overflow dataset (22.10\% average). Developers frequently seek explanations for issues and solutions to real-time problems, highlighting the need for robust documentation and support mechanisms.
\newline
\textbf{Feature Exploration and Innovation}: The significant presence of Feature Requests and Improvements in the OpenAI dataset (e.g., 37.9\% in Web and App Integration) complements the exploratory ‘What’ questions from Stack Overflow (16.41\%). This suggests developers’ active interest in expanding capabilities and exploring new features for LLMs.
\newline
\textbf{Guidance and Best Practices}: The OpenAI dataset’s Guidance and Validation category aligns with ‘Why’ questions from Stack Overflow, showing developers’ need for validation, advice, and deeper insights into their workflows. Ethical and best practices discussions in both datasets remain niche but underscore their importance in specialized contexts.
\newline
\textbf{Emerging Trends}: Unique to the OpenAI dataset are categories like Maintenance and Ongoing Support, which, while less represented, highlight long-term project management considerations. These categories are less evident in Stack Overflow, potentially reflecting differences in the scope and focus of discussions on each platform.
\end{boxK}

\subsection{RQ3: To what extent do developers perceive the revealed challenges in terms of difficulties?}
\textbf{Motivation: }Building on our understanding of various topics identified from Stack Overflow and OpenAI Developer Forum, our next step is to assess the difficulty level of each topic. Determining which topics are most challenging to address on these platforms is crucial for identifying where developers encounter the most significant obstacles. As LLM technology adoption accelerates, developers face increasingly complex technical issues that demand specialized support and expertise. By pinpointing difficult topics, we can gain valuable insights into gaps and limitations in current documentation and support resources. This analysis will help tailor educational resources to areas of greatest need, benefiting the broader developer community by revealing opportunities to enhance accessibility and support for complex issues.
\newline
\textbf{Approach: }We assess the difficulty of each topic using a range of metrics. For Stack Overflow, we consider the following metrics to evaluate the difficulty level of each topic:
\newline
\textbf{Metrics for Stack Overflow:}
\begin{itemize}
    \item \textbf{The percentage of posts of a topic without accepted answers (\% w/o accepted answers). }For each LLM topic obtained from Stack Overflow, we assess the difficulty by measuring the percentage of posts lacking accepted answers. While a post may receive multiple responses, only the original author can designate one as accepted if it sufficiently addresses their question. Therefore, topics with fewer accepted answers are considered more challenging \cite{rosen2016mobile, abdellatif2020challenges, bagherzadeh2019going}.
    \item \textbf{The median time for an answer to be accepted (median time to answer (hrs.)).} We measure the median time in hours for posts to receive an accepted answer, basing this metric on when the answer is created rather than when it is marked as accepted. A longer duration to receive an accepted answer suggests that the post poses more significant challenges \cite{abdellatif2020challenges, rosen2016mobile, bagherzadeh2019going}.
\end{itemize}

To explain the difficulty level of challenges of OpenAI developer forum data faced by LLM developers, we analyze the number of replies to questions related to LLM development. We focus on analyzing the number of replies to LLM development-related questions, assuming that a higher number of replies suggests that the topic is easier to address, and, in contrast, questions with fewer replies typically indicate more challenging issues \cite{chen2024empirical}.

\begin{table}[htbp]
    \centering
    \scriptsize
    \caption{Difficulty Metrics for LLM-Related Topics on Stack Overflow}
    \label{tab:llm_topic_difficulty}
    \begin{tabular}{p{7cm}|r|r}
        \toprule
        \textbf{Topic} & \textbf{Posts w/o Accepted (\%)} & \textbf{Median Time (h)} \\
        \midrule
        LLM Ecosystem and Challenges & 82.20 & 21.57 \\
        API Usage & 74.99 & 10.53 \\
        Environment Management & 74.46 & 26.91 \\
        LLM Training with Frameworks & 63.41 & 14.10 \\
        Programming Constructs and LLM Integration & 81.03 & 45.22 \\
        Llama Indexing and GPU Utilization & 85.07 & 55.76 \\
        Audio Transcription and Speech Recognition Automation & 79.53 & 44.12 \\
        Langchain Development and Error Handling & 75.36 & 25.10 \\
        Agents and Tool Interactions & 90.63 & 36.51 \\
        \bottomrule
    \end{tabular}
\end{table}

The table \ref{tab:llm_topic_difficulty} provides an analysis of the difficulty associated with various LLM-related topics on Stack Overflow, using two key metrics: the percentage of posts without an accepted answer and the median response time (in hours). 

The topic \emph{Agents and Tool Interactions} demonstrates the highest level of difficulty, evidenced by the highest proportion of unanswered posts (90.63\%) and a substantial median response time of 36.51 hours. These values suggest that questions in this domain may involve complex interactions or require highly specialized knowledge, contributing to a lower resolution rate. Similarly, \emph{Llama Indexing and GPU Utilization} exhibits significant difficulty, with 85.07\% of posts lacking accepted answers and the longest median response time of 55.76 hours. This indicates a high degree of complexity or niche expertise that may limit community engagement and effective problem-solving.

Topics such as \emph{LLM Ecosystem and Challenges} and \emph{Programming Constructs and LLM Integration} also present notable difficulty levels, with 82.20\% and 81.03\% of posts remaining unresolved, and median response times of 21.57 and 45.22 hours, respectively. These metrics highlight potential gaps in community expertise or the intricate nature of issues within these areas. In contrast, \emph{LLM Training with Frameworks} and \emph{API Usage} appear less challenging. These topics have lower percentages of unresolved posts (63.41\% and 74.99\%, respectively) and comparatively shorter median response times (14.10 and 10.53 hours), indicating that they may be more familiar to contributors or involve more straightforward technical problems.

\begin{boxK}
    Overall, the analysis suggests that topics involving more specialized, intricate, or interdisciplinary aspects of LLMs, such as agent-based interactions and GPU utilization, pose greater challenges for the community. These challenges are reflected in the higher proportions of unresolved posts and extended response times, indicating a potential need for more targeted expertise and resources in these areas.
\end{boxK}
\begin{table}[htbp]
    \centering
    \scriptsize
    \caption{Correlation of topics' popularity and difficulty metrics.}
    \label{tab:correlation_popularity_difficulty}
    \begin{tabular}{p{3.5cm}|r|r}
        \toprule
        Correlation Coeff./p-value & \% w/o Accepted Ans & Median Time(H) \\
        \midrule
        \textbf{Posts} & 0.070/0.858 & -0.423/0.256 \\
        \textbf{Avg. Views} & -0.862/0.003 & -0.706/0.034 \\
        \textbf{Avg. Score} & -0.829/0.006 & -0.524/0.148 \\
        \bottomrule
    \end{tabular}
\end{table}

To better understand the characteristics of LLM-related posts, we investigate the correlation between their difficulty and popularity. We employ the Spearman Rank Correlation Coefficient \cite{ref1} to evaluate the relationships between two popularity metrics (average views and average scores) and two difficulty metrics (percentage of posts without accepted answers and median response time in hours). This approach allows us to determine whether these popularity and difficulty measures are statistically related, providing insights into how community engagement and response challenges align.

Table \ref{tab:correlation_popularity_difficulty} presents the correlation coefficients and p-values obtained from analyzing the relationships between the popularity and difficulty metrics of LLM-related topics on Stack Overflow.The results indicate no significant correlation between the \emph{number of posts} and difficulty metrics. Specifically, the correlation between the number of posts and the percentage of posts without an accepted answer is weak (0.070) with a high p-value (0.858), indicating no meaningful relationship. Additionally, the correlation with median response time is negative (-0.423) but also not statistically significant (p-value of 0.256).

In contrast, \emph{average views} demonstrate a strong and statistically significant negative correlation with the difficulty metrics. The correlation coefficient between average views and the percentage of unanswered posts is -0.862 (p-value = 0.003), suggesting that topics with higher visibility tend to have a lower proportion of unanswered posts. Similarly, the correlation between average views and median response time is -0.706 (p-value = 0.034), indicating that more frequently viewed topics are typically resolved more quickly. The \emph{average score} metric also correlates negatively with the difficulty metrics. The correlation between the average score and the percentage of unanswered posts is -0.829 (p-value = 0.006), showing a significant relationship where higher-scoring topics are less likely to remain unresolved. Although the correlation between average score and median response time is moderate (-0.524) with a p-value of 0.148, suggesting a trend toward quicker resolution for higher-scoring posts, it is not statistically significant.
\begin{table*}[htbp]
\centering
   \caption{Difficulty Metrics for LLM-Related Topics on OpenAI Developer Forum}
\label{tab:reply_range_counts}
\begin{adjustbox}{max width=\textwidth}
\begin{tabular}{lrrrrrr}
\toprule
\textbf{Topic} & \textbf{0 Replies} & \textbf{1-2 Replies} & \textbf{3-5 Replies} & \textbf{6-10 Replies} & \textbf{10+ Replies} & \textbf{Total} \\
\midrule
API Usage and Error Handling & 1940 & 3470 & 2510 & 1390 & 903 & 10213 \\
LLM Functionalities & 723 & 1507 & 1159 & 706 & 578 & 4673 \\
Managing API Requests and Responses & 407 & 706 & 539 & 269 & 166 & 2087 \\
Integrating ChatGPT into Web and App Services & 282 & 495 & 336 & 196 & 168 & 1477 \\
AGI and AI Project Collaboration & 261 & 396 & 263 & 138 & 97 & 1155 \\
Chat Completion and Bot Development & 184 & 369 & 284 & 152 & 82 & 1071 \\
Intelligent Agents and Generative Models & 222 & 295 & 195 & 112 & 103 & 927 \\
Token Usage & 92 & 252 & 226 & 121 & 70 & 761 \\
Fine-Tuning and Dataset Management & 123 & 249 & 162 & 126 & 77 & 737 \\
File Management and Retrieval & 141 & 186 & 129 & 72 & 47 & 575 \\
Prompt Engineering & 71 & 149 & 126 & 87 & 58 & 491 \\
ChatGPT Models & 64 & 132 & 103 & 68 & 50 & 417 \\
Plugin Development & 69 & 116 & 115 & 63 & 47 & 410 \\
Image Generation & 85 & 129 & 103 & 50 & 37 & 404 \\
Embedding & 42 & 116 & 115 & 63 & 48 & 384 \\
Data Preparation and Structured Analysis & 63 & 132 & 90 & 42 & 20 & 347 \\
Function Parameters and Callback Handling & 75 & 102 & 79 & 60 & 29 & 345 \\
\bottomrule
\end{tabular}
\end{adjustbox}
\end{table*}

\begin{boxK}
    Overall, the results indicate that the popularity of topics, as measured by average views and scores, is inversely related to their difficulty. Topics that garner more attention and higher scores tend to have fewer unresolved posts and shorter response times, underscoring the importance of community engagement in facilitating prompt and effective answers.
\end{boxK}

The analysis of reply count distribution reveals distinct patterns of community engagement and topic difficulty within the OpenAI Developer Forum. Table \ref{tab:reply_range_counts} shows the reply count distribution of the generated topics. Topics such as \emph{API Usage and Error Handling} and \emph{LLM Functionalities} exhibit high numbers of posts across all reply ranges, including a significant number of unanswered questions (i.e., reply range '0'). This suggests that while these topics attract substantial community attention, they also contain a number of unresolved challenges, indicating their inherent complexity. In contrast, topics like \emph{Managing API Requests and Responses} and \emph{Integrating ChatGPT into Web and App Services} have moderate engagement with fewer posts in the '0' reply range, implying that these areas are comparatively easier or that the community is better equipped to address questions. On the other hand, smaller topics such as \emph{Prompt Engineering}, \emph{ChatGPT Models}, and \emph{Function Parameters and Callback Handling} have fewer posts overall, with many falling in the '1-2' reply range, suggesting moderate engagement but a smaller active community. Generally, the topics with higher numbers of posts without replies—particularly those involving API usage and GPT models—highlight gaps in community knowledge or complexities that require additional resources or expertise to resolve effectively. This suggests that these topics might benefit from further exploration, better documentation, or specialized assistance to reduce their difficulty for developers.
\begin{boxK}
    \textbf{RQ3 Summary: }The combined analysis of difficulty metrics for LLM-related topics from the OpenAI Developer Forum and Stack Overflow reveals distinct areas of challenge for developers. \emph{API Usage and Error Handling} consistently emerges as one of the most difficult topics across both platforms, characterized by a high number of posts with zero replies (1,940) on the OpenAI forum and a 74.99\% rate of posts without accepted answers on Stack Overflow. Similarly, \emph{Agents and Tool Interactions} show significant difficulty, with 90.63\% of Stack Overflow posts unresolved and notable engagement challenges in the forum data. \emph{Llama Indexing and GPU Utilization} also stands out for its complexity, with 85.07\% of Stack Overflow posts lacking accepted answers and a lengthy median response time of 55.76 hours. Other challenging areas include \emph{LLM Ecosystem and Challenges}, \emph{Programming Constructs and LLM Integration}, and \emph{Audio Transcription and Speech Recognition Automation}, all marked by high percentages of unresolved posts and substantial delays in responses.  
    Conversely, topics like \emph{Prompt Engineering}, \emph{ChatGPT Models}, and \emph{LLM Training with Frameworks} exhibit comparatively lower difficulty levels, with fewer unresolved posts and shorter response times. Overall, the data suggests that integration, indexing, and complex agent usage are areas needing additional support, while general LLM training and prompt-related topics are more approachable for developers. Addressing these difficulties could enhance the developer experience and facilitate smoother adoption of LLM technologies.
\end{boxK}

\section{Discussion \& Implications of our Findings}\label{discussion-implications}
In this section, we explore the evolution of LLM-related topics and compare our findings with those of prior studies. We then analyze the data to identify the dominant topics across different platforms and discuss the implications of our findings.
\begin{figure}[htbp]
\centering
  \vspace{-0.8em}
  \includegraphics[width=\textwidth, height=0.3\textheight]{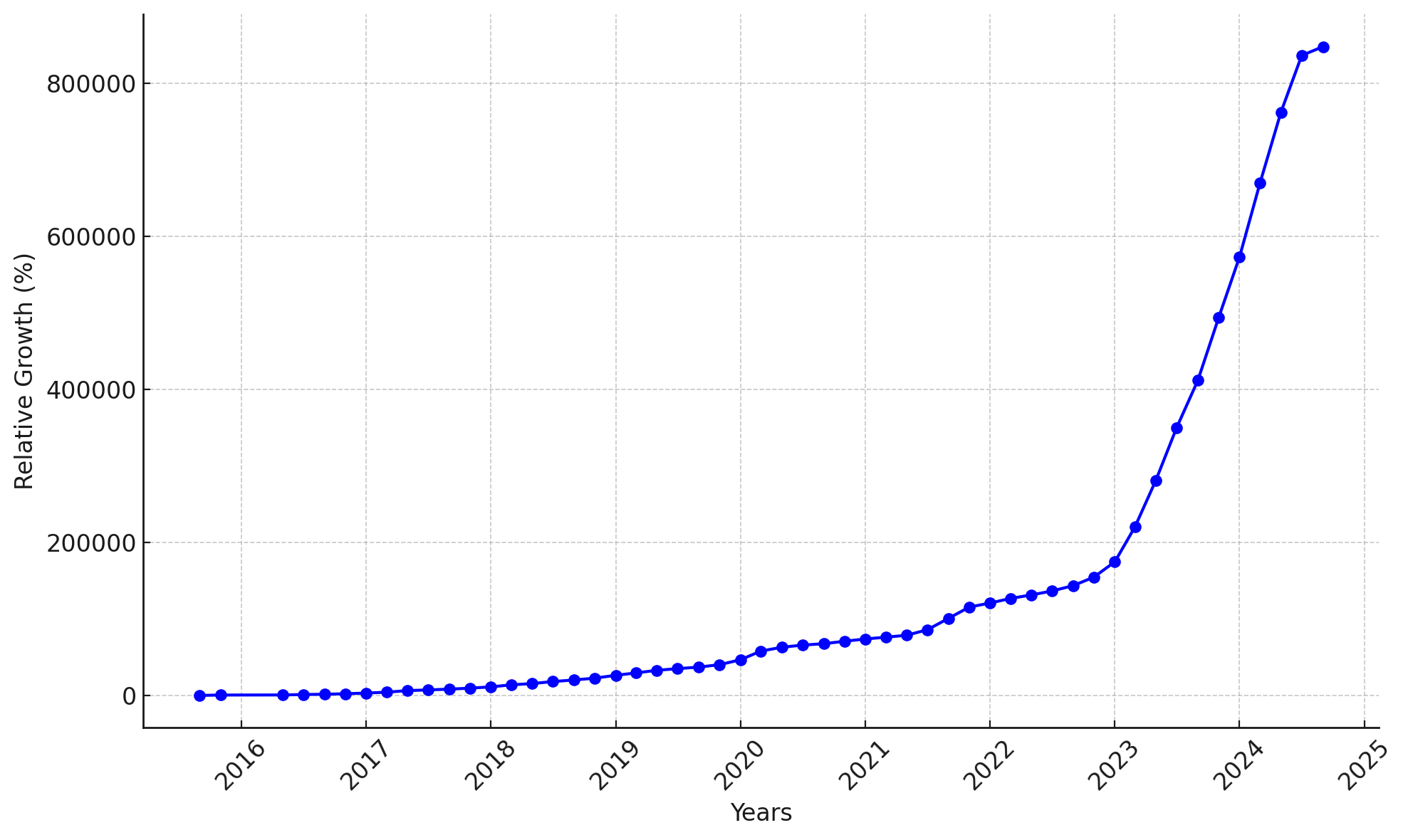}
  \caption{Relative Growth of LLM-related posts over time in Stack Overflow}
  \label{fig:llmgrowth}
  \vspace{-1 em}
\end{figure}
\subsection{LLM Topics Evolution}
LLMs are rapidly emerging as a focal point for developers across diverse fields, including databases \cite{li2024can}, chatbots \cite{kim2023chatgpt}, and software engineering \cite{hou2023large}. Figure \ref{fig:llmgrowth} illustrates a clear upward trend in developer engagement with LLM topics over time. Starting modestly, the volume of related posts has grown steadily, with a marked increase in recent years. This surge likely aligns with key advancements, such as the releases of major LLMs like GPT-3, GPT-4, and others, which have broadened LLM applicability and spurred interest. The sustained growth in posts reflects the expanding impact of LLMs across industries as developers explore their applications, address challenges, and deepen their integration into various domains.

\begin{figure}[htbp]
\centering
  \vspace{-1em}
  \includegraphics[width=\textwidth]{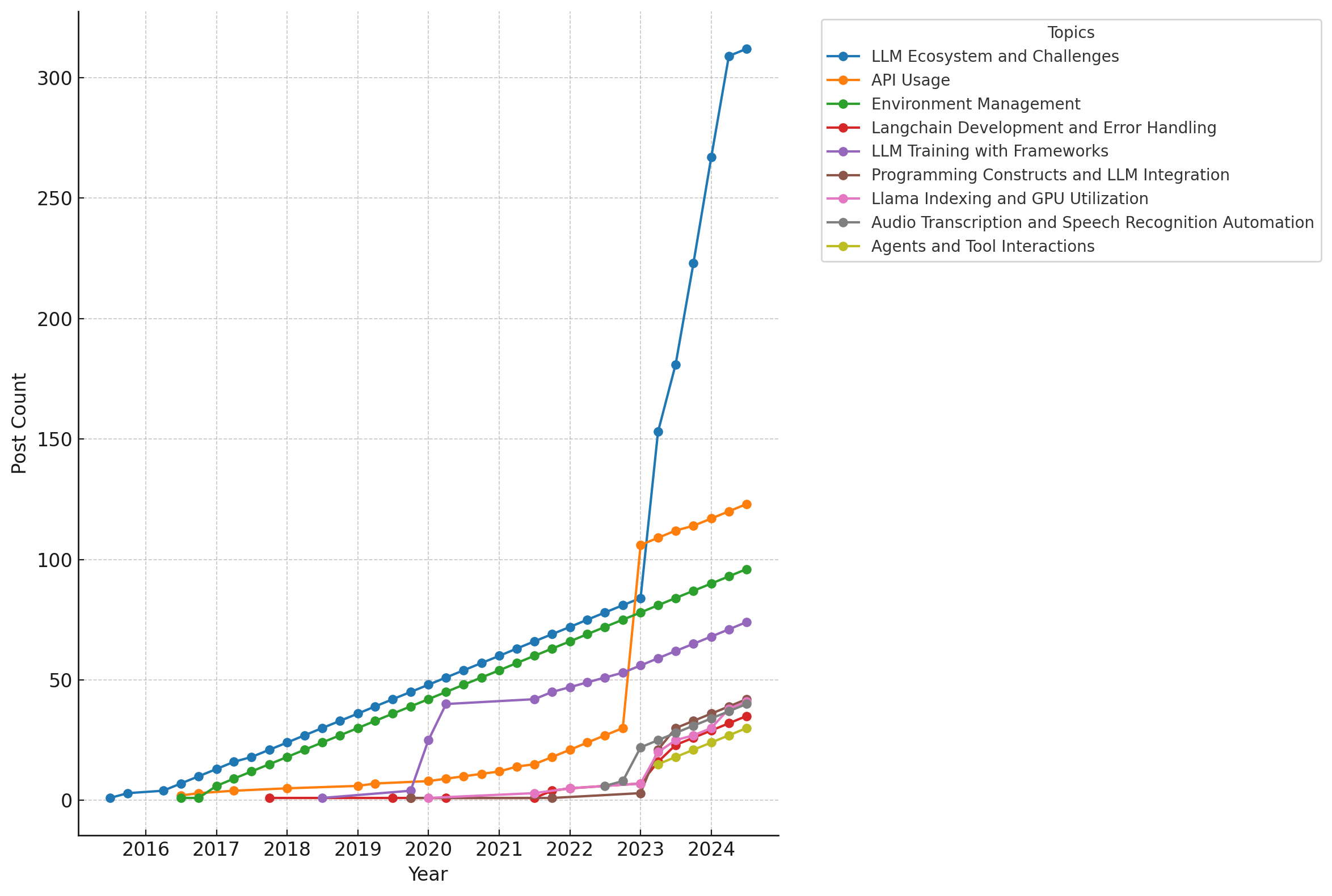}
  \caption{LLM Topic Evolution Over Time in Stack Overflow}
  \label{fig:topicwisedata}
  \vspace{-0.8em}
\end{figure}
To better understand the evolution of various LLM development activities, we present the topic-wise progression of each area in figure \ref{fig:topicwisedata}. The figure reveals a dynamic shift in interest across LLM-related topics on Stack Overflow, with some topics showing steady growth and others experiencing sharper, periodic spikes. Areas, such as \emph{LLM Ecosystem and Challenges} and \emph{API Usage}, demonstrate consistent upward trends, reflecting sustained developer interest and ongoing challenges. In contrast, topics like \emph{LLM Training with Frameworks} and \emph{Llama Indexing and GPU Utilization} show rapid increases, likely driven by recent advancements and rising computational demands. Topics like \emph{Audio Transcription and Speech Recognition Automation} display moderate, stable growth, indicating niche but persistent engagement. This variation in growth patterns—steady for core topics and intermittent for specialized areas—highlights the complexity of the LLM field and points to both broad and targeted interest among developers.
\begin{figure}[htbp]
    \centering
  \vspace{-1em}
  \includegraphics[width=\textwidth]{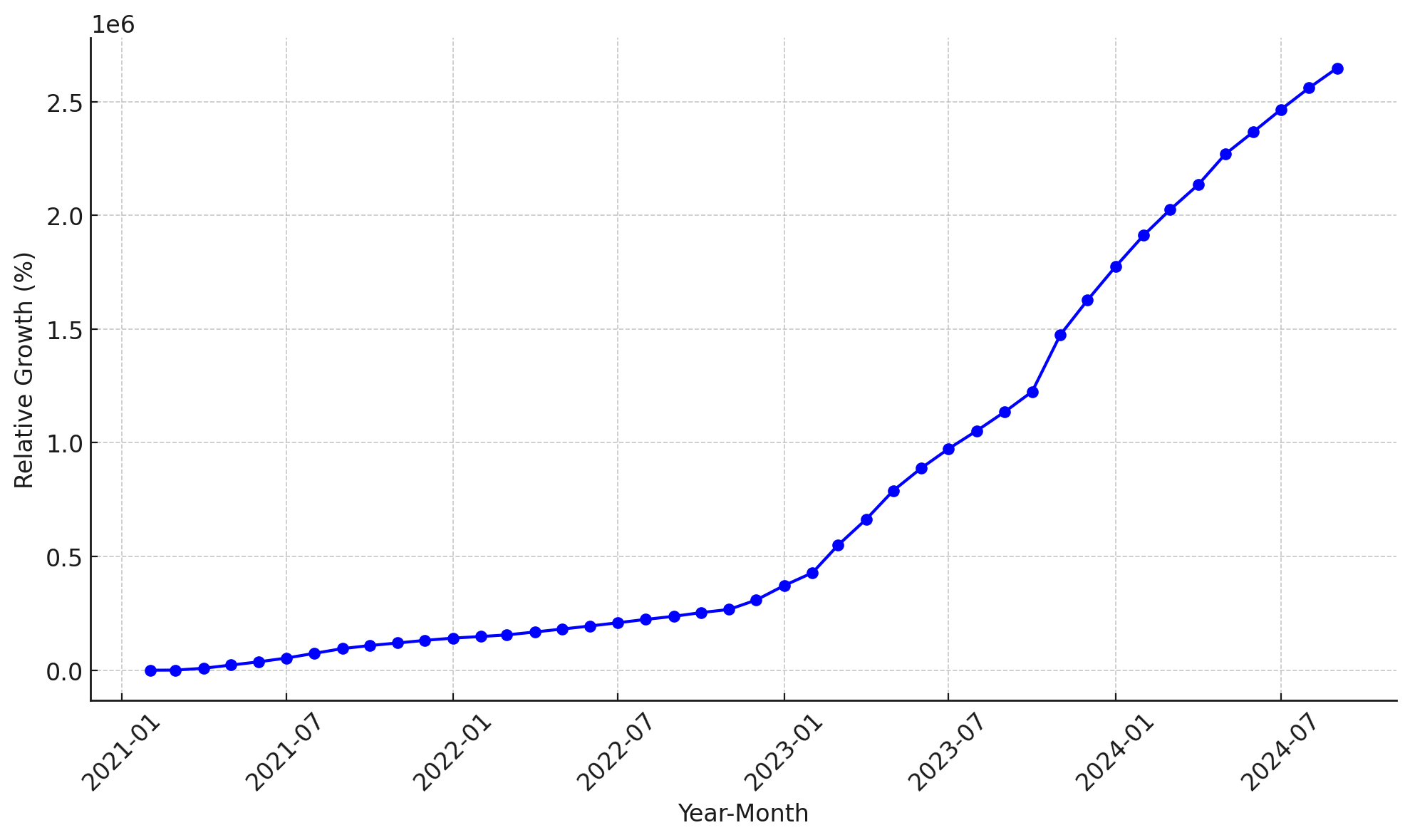}
  \caption{Relative Growth of LLM-related posts over time in OpenAI Developer Forum}
  \label{fig:llmgrowthopenai}
  \vspace{-0.8em}
\end{figure}
The relative growth plot \ref{fig:llmgrowthopenai} of posts on the OpenAI Developer Forum illustrates the rapid evolution of developer interest in LLMs. Initially, there was modest growth, reflecting early engagement as foundational LLM tools gained traction. However, as advanced models like GPT-3 and GPT-4 were introduced, relative growth surged sharply, indicating heightened interest and experimentation. This spike aligns with the expansion of LLM applications across diverse fields, where developers face new challenges and opportunities. This pattern reflects a typical trajectory of transformative technology, evolving from early adoption to broader, focused integration.
\begin{figure}[htbp]
\centering
  \vspace{-2em}
  \includegraphics[width=\textwidth]{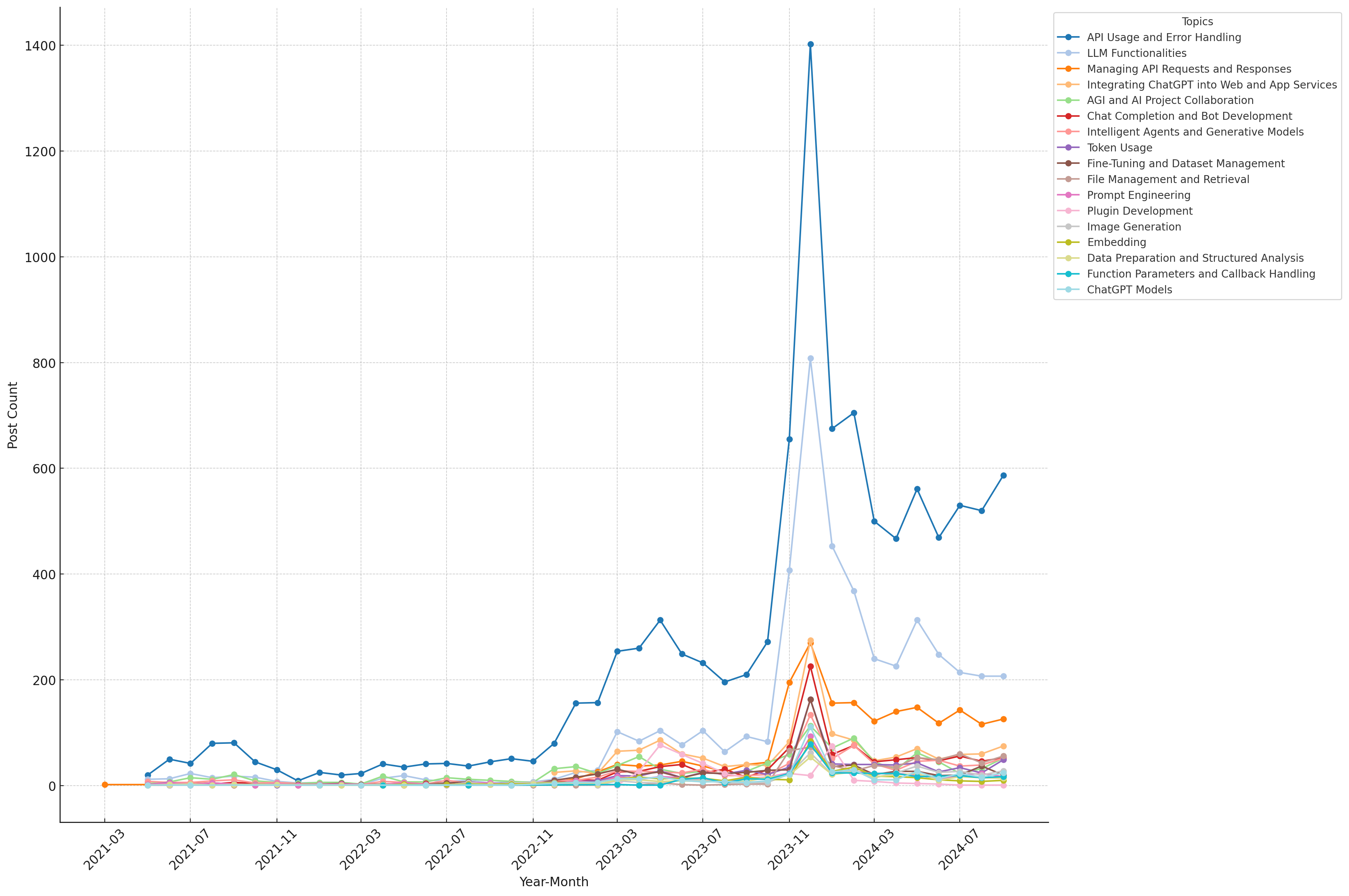}
  \caption{LLM Topic Evolution Over Time in OpenAI Developer Forum}
  \label{fig:topicevolutionopenaidata}
  \vspace{-0.8em}
\end{figure}

We also explore the evolution of each topic to understand the broader trends driving their growth over time. Figure \ref{fig:topicevolutionopenaidata} illustrates the evolution of each topic, focusing on LLM-related discussions from 2021 to 2024. The progression of these topics reveals a substantial transformation in the level of community interest and engagement. Initially, during 2021 and early 2022, activity around LLMs was modest, as evidenced by the low post counts. This suggests that LLM technologies were primarily confined to experimental or niche research settings at that time. However, beginning in late 2023, there was a sharp surge in activity that continued into 2024, reflecting the widespread adoption of LLM technologies and a growing interest from the developer community. This surge can be attributed to significant advancements in LLM models, increased accessibility of these technologies, and the emergence of practical use cases that demonstrated their potential utility.

The rapid growth in activity was most prominent in topics related to LLM ecosystems, training methodologies, API integration, and implementation, highlighting the increasing focus of developers on understanding and deploying LLMs effectively in real-world scenarios. This trend highlights how developers progressively tackle complex challenges such as training, fine-tuning, and effectively integrating LLMs into operational systems. The evolution of these topics underscores the transition of LLMs from specialized research tools to mainstream technologies, embraced by an active and expanding community.

\subsection{LLM Compared to Other Software Engineering Fields}
In the previous sections, we find a significant rise in discussions surrounding LLMs, beginning around 2020 and accelerating notably in late 2022. This increase aligns with the release of advanced LLM-based platforms like ChatGPT, powered by GPT, and Gemini, underscoring LLMs as an emerging and rapidly advancing field. As a new and emerging field, we set out to investigate how the topics of LLM compare against those of other Software Engineering fields, such as Chatbot Development, Quantum Software Engineering, Mobile Applications, Big Data, and Security—previously examined in various studies. 

\begin{table*}[htbp]
    \centering
    \caption{Comparison of Popularity and Difficulty Between Different Fields}
    \label{tab:differentfieldcompare}
    \begin{adjustbox}{max width=\textwidth}
    \begin{tabular}{lrrrrrrr}
        \toprule
        \textbf{Metrics} & \textbf{LLM} & \textbf{Chatbot} & \textbf{QSE} & \textbf{WebApp} & \textbf{Mobile} & \textbf{Security} & \textbf{BigData} \\
        \midrule
        \# of Posts & 8,593 & 7,295 & 722 & 14,596 & 2,502,811 & 341,180 & 197,482 \\
        Avg ViewCount & 2,063.93 & 1,013.78 & 717.03 & 2,469.17 & 2,792.04 & 3,436.26 & 2,795.66 \\
        Avg Score & 1.25 & 0.77 & 1.08 & 2.08 & 2.39 & 2.66 & 1.75 \\
        Avg AnswerCount & 0.87 & 1.05 & 1.04 & 1.00 & 1.49 & 1.39 & 1.18 \\
        \% w/o Accepted Answers & 79.03\% & 68.17\% & 62.88\% & 64.92\% & 53.17\% & 52.34\% & 59.07\% \\
        Med. TimeToAns (Hrs) & 19.88 & 15.94 & 9.59 & 11.41 & 0.93 & 1.89 & 3.29 \\
        \bottomrule
    \end{tabular}
    \end{adjustbox}
\end{table*}

To achieve this, we examine the popularity and difficulty of LLM-related topics and compare these metrics with those from other fields. The analysis incorporates data from Stack Overflow, drawing on insights from studies focusing on Chatbot challenges \cite{abdellatif2020challenges}, Quantum Software Engineering \cite{li2021understanding}, Mobile Applications \cite{rosen2016mobile}, Big Data \cite{bagherzadeh2019going}, and Security \cite{yang2016security}. These past studies span different periods, so to ensure comparability, we follow the methodology outlined by Abdellatif et al. \cite{abdellatif2020challenges}. This approach uses reported keywords from each study to reconstruct and expand the dataset, facilitating a consistent framework for measuring popularity and difficulty across fields.

Table \ref{tab:differentfieldcompare} shows that while LLM and Chatbot development have gained traction only in recent years, long-established domains like Mobile, Security, and WebApp development benefit from over a decade of accumulated community knowledge, resulting in faster response times and a higher rate of accepted answers. Mobile development, for instance, sees rapid responses (median of 0.93 hours) and fewer unanswered questions (53.17\%), reflecting its well-supported community. Conversely, LLM, though relatively new, already has significant engagement with 8,593 posts and an average view count comparable to mature fields. However, its high percentage of questions without accepted answers (79.03\%) and lengthy median time to answer (19.88 hours) highlight the complexity and novelty of this field, where fewer established practices and a smaller pool of experts make answering questions more challenging. This contrast underscores the maturity gap: established fields benefit from sustained community support and a wealth of best practices, while LLM, despite rapid growth, remains a challenging field due to limited expertise and evolving methodologies.

\subsection{Implications}
The results of our study can help the LLM community focus more effectively on the most pressing issues in LLM development. Based on the popularity and difficulty data for LLM-related topics provided by Stack Overflow and OpenAI Developer Forum in the previous sections, below, we describe how these findings can guide practitioners, researchers, and educators in advancing the practice and learning of LLM development.
\newline
\textbf{Implication for Researchers: }The findings reveal specific LLM topics that are popular yet challenging to resolve, indicating areas where further research is needed. For example, \emph{LLM Ecosystem and Challenges} and \emph{Programming Constructs and LLM Integration} have high difficulty metrics, with long resolution times and high percentages of unresolved posts. This suggests gaps in current knowledge and tools, making these topics prime candidates for future research. Researchers could investigate improved frameworks, debugging tools, or systematic studies to reduce difficulty in these areas. Additionally, high-interest areas like \emph{API Usage and Error Handling} and \emph{Integrating ChatGPT into Web and App Services} indicate a need for more robust solutions and better guidance on effective API management and integration strategies.
\newline
\textbf{Implication for Practitioners: }For practitioners, especially developers and engineers, this study highlights specific technical areas where they will likely face difficulty. Topics like \emph{Llama Indexing and GPU Utilization} and \emph{Audio Transcription and Speech Recognition Automation} show high difficulty but relatively lower popularity, suggesting they require specialized skills. Practitioners could benefit from focused training or additional resources in these areas to avoid roadblocks in development. Furthermore, the high popularity and moderate difficulty of topics like \emph{API Usage} and \emph{Chat Completion and Bot Development} point to common challenges where standard best practices and troubleshooting guides could reduce repeated errors and improve efficiency.
\newline
\textbf{Implication for Educators: }Educators can use these findings to design curricula that prepare students for real-world challenges in LLM development. Topics with high popularity and difficulty, such as \emph{LLM Ecosystem and Challenges} and \emph{Langchain Development and Error Handling}, could be covered through case studies, hands-on labs, and group projects, allowing students to tackle these complex issues in a controlled learning environment. Additionally, topics like \emph{Token Usage} and \emph{Prompt Engineering}, which have high popularity on the OpenAI Developer Forum, indicate emerging skills that would be valuable for students entering the industry. By addressing fundamental and challenging areas in their curricula, educators can better equip students with the knowledge and skills they need to succeed in LLM-related developments.

Practitioners, researchers, and educators have many factors to consider when choosing where to focus their efforts. However, we believe that our findings and insights can meaningfully support and guide this decision-making process.

\section{Threats to Validity} \label{threats-to-validity}
Threats to validity are factors that can compromise the accuracy, reliability, or generalizability of research findings. They pertain to the potential disconnect between study conclusions and actual reality \cite{bean2007qualitative}. Such threats can introduce errors, biases, or limitations that compromise the validity and credibility of the results. Recognizing and addressing these threats is essential to enhance the robustness and trustworthiness of research outcomes. In our study, we acknowledge the following potential threats.

\textbf{Internal Validity: }refers to factors that could impact the accuracy of our results \cite{wohlin2012experimentation}. In this study, we identify LLM-related posts on Stack Overflow by using tags, acknowledging that some relevant posts might be misclassified due to missing or incorrect tags. To mitigate this, we examine tags co-occurring with the 'large-language-models' tag and used TST and TRT measures to select related tags. Similar measures have been used in previous research \cite{abdellatif2020challenges, rosen2016mobile, wan2019programmers, yang2016security}, to enhance better coverage while minimizing noise in the dataset. Our TST and TRT thresholds align with those established in prior studies, further supporting our methodology. In addition to our primary tag set, we also include the 'openai' tag for data extraction, as OpenAI plays a pivotal role in the development of LLMs. 

For topic modeling, we employ BERTopic to cluster similar posts from Stack Overflow and OpenAI Developer Forum based on the assumption that similar content would form similar clusters. However, alternative methods could result in different clusters of posts. To ensure the quality of the clusters, we manually review the generated topics and assign meaningful labels to each cluster. Another potential concern is selecting the optimal number of topics. While BERTopic provides a 'nr\_topics' parameter for topic control, we opt to determine the number of topics through clustering using the HDBSCAN model. To enhance data representation, we utilize CountVectorizer, which allows us to filter out infrequent words and expand the n-gram range, thereby improving the relevance and granularity of the topics generated.

The labeling process for Stack Overflow and OpenAI Developer Forum posts introduces some subjectivity, which we address through independent classifications and interrater reliability testing. Using Cohen's Kappa, we confirm strong agreement among annotators on Stack Overflow posts. For OpenAI Developer Forum data, we leverage ChatGPT-4 and our combined expertise, enhancing the consistency and reliability of our classifications. Together, these steps collectively enhance the internal Validity of our findings, supporting their robustness and credibility.

\textbf{Construct Validity: }addresses the relationship between theoretical concepts and observed data \cite{wohlin2012experimentation}. A potential concern in our study is the accuracy of labeling automatically generated topics, as the assigned names may not fully capture the essence of the related posts. To mitigate this, three authors independently review the keywords and examine over 30 randomly selected posts for each topic. We then discussed and reached a consensus on labels that best represented the content of each topic.

We also assess the difficulty of LLM topics using metrics such as the percentage of unresolved posts, reply count, and median time to resolution. While these metrics provide useful insights, they could impact construct validity. Our choice of metrics is consistent with prior research \cite{chen2024empirical, abdellatif2020challenges, ahmed2018concurrency, bajaj2014mining, nadi2016jumping, rosen2016mobile, yang2016security}, reinforcing their reliability for measuring topic difficulty in developer forums.

\textbf{External Validity: }refers to the extent to which research findings can be generalized to broader populations or contexts \cite{wohlin2012experimentation, ferguson2004external}. In this study, we analyze data from Stack Overflow and OpenAI Developer Forum to gain insights into the challenges faced by LLM developers. Initially, we rely on Stack Overflow to identify the challenges, though this source may not capture every nuance of developers’ difficulties, as some issues are less commonly discussed. To address this limitation, we also examine OpenAI Developer Forum data, capturing additional challenges in LLM development. Nonetheless, developers may discuss challenges on other platforms, such as mailing lists and other LLM forums, which could provide further insights.

Our focus on Stack Overflow and OpenAI Developer Forum enhances the generalizability of our findings. Stack Overflow is a widely recognized platform with a diverse user base covering multiple domains and expertise levels, while OpenAI Developer Forum has a nearly one-million-strong community with varied skill levels. Together, these platforms provide a comprehensive view of challenges in LLM development. However, future research could improve upon this study by incorporating data from other discussion platforms or directly surveying LLM developers to capture a broader range of experiences. Expanding the data sources would strengthen the generalizability and depth of our findings and offer a more holistic view of the challenges in LLM development.

\section{Conclusion} \label{conclusion}
In this study, we shed light on developers' challenges when working with Large Language Models by analyzing community interactions on Stack Overflow and OpenAI Developer Forum. Our analysis identifies nine LLM-related challenges on Stack Overflow and 17 on the OpenAI Developer Forum, highlighting key areas of difficulty, including implementation, API usage, fine-tuning, and model integration. We also find that developers encounter traditional software engineering challenges, such as API Usage and Error Handling and Environment Management, as well as unique challenges specific to LLMs, including LLM Training with Frameworks, Agents, and Tool Interactions, Fine-Tuning, and Integrating models with other applications. Our findings reveal that certain LLM areas, such as API usage and error handling, LLM functionalities, and the LLM ecosystem and challenges, attract significant attention from developers while posing substantial difficulties. Our findings also disclose that LLM-related queries are generally more complex and have slower resolution rates than more established domains, such as mobile and security. The high number of unresolved posts and prolonged response times for complex LLM-specific issues, particularly those involving GPU utilization and agent interactions, underscore the need for specialized support and resources tailored to these challenges. 

This research offers valuable insights for the LLM community by identifying the most challenging aspects of LLM development, which require targeted research and resource allocation. Our findings suggest a growing demand for improved tools, documentation, and community support to address the complex, evolving needs of LLM practitioners. Future work could focus on expanding resources, developing automated support tools, and promoting enhanced community engagement to meet the evolving needs of LLM developers. This study can serve as a foundation for further research that will ultimately strengthen and support LLM development and deployment across diverse applications, contributing to a more robust and accessible ecosystem for LLM innovation.

\section*{Acknowledgment}
This research is supported in part by the Natural Sciences and Engineering Research Council of Canada (NSERC), and by the industry-stream NSERC CREATE in Software Analytics Research (SOAR).
\appendix

 \bibliographystyle{elsarticle-num} 
 \bibliography{elsarticle-template-num}

\end{document}